\documentclass[twocolumn,prc,aps,showpacs,preprintnumbers,amsmath,amssymb]{revtex4}

\usepackage{graphicx}

 \def\Pom{{ I\!\!P}}
 \def\Reg{{ I\!\!R}}
 \def\gsim{\mathrel{\rlap{\lower4pt\hbox{\hskip1pt$\sim$}}
 \raise1pt\hbox{$>$}}}

 \newcommand\la{\langle}
 \newcommand\ra{\rangle}
 \newcommand\beq{\begin{equation}}
 \newcommand\noi{\noindent}
 \newcommand\eeq{\end{equation}}
 \newcommand\beqn{\begin{eqnarray}}
 \newcommand\eeqn{\end{eqnarray}}
\def\mb{\,\mbox{mb}}
\def\fm{\,\mbox{fm}}
\def\GeV{\,\mbox{GeV}}
\def\TeV{\,\mbox{TeV}}
\def\lsim{\mathrel{\rlap{\lower4pt\hbox{\hskip1pt$\sim$}}
    \raise1pt\hbox{$<$}}}         
\def\gsim{\mathrel{\rlap{\lower4pt\hbox{\hskip1pt$\sim$}}
    \raise1pt\hbox{$>$}}}         
\def\Re{\,\mbox{Re}\,}
\def\Im{\,\mbox{Im}\,}
\def\mb{\,\mbox{mb}}
\def\fm{\,\mbox{fm}}  
\def\GeV{\,\mbox{GeV}}

\def\T{T^h_A(b)}
\def\s0{\sigma_0(s)}
\def\sq{\sigma_{\bar qq}}
\def\st{\sigma_{tot}^{pN}}
\def\sel{\sigma_{el}^{pN}}
\def\sinhad{\sigma_{in}^{pN}}

\def\sqa{\sigma_{\bar qq}^A}
\def\sta{\sigma_{tot}^{pA}}
\def\sela{\sigma_{el}^{pA}}

\def\stn{\sigma_{tot}^{NN}}
\def\seln{\sigma_{el}^{NN}}

\def\doublespace{\def\baselinestretch{1.6}\large\normalsize}
\def\normalspace{\def\baselinestretch{1.0}\normalsize}
\def\Caption#1{
  \normalspace
  \begin{quotation}\caption{\sl #1}\end{quotation}
  \doublespace
}
\begin{document}
\date{}

\title{\bf Large Rapidity Gap Processes\\ in Proton-Nucleus Collisions}
 
 
 
\author{B.Z.~Kopeliovich$^{1,2}$} 
\author{I.K.~Potashnikova$^{1,2}$} 
\author{Ivan~Schmidt$^1$}

\affiliation{$^1$ Departamento de F\'isica, Universidad Federico Santa 
Mar\'ia, Valpara\'iso, Chile\\
$^2$ Joint Institute for Nuclear Research, Dubna, Russia}

 

\date{\today}

\begin{abstract} 

The cross sections for a variety of channels of proton-nucleus interaction
associated with large gaps in rapidity are calculated within the Glauber-Gribov
theory. We found inelastic shadowing corrections to be dramatically enhanced
for such events. We employ the light-cone dipole formalism which allows to
calculate the inelastic corrections to all orders of the multiple interaction.
Although Gribov corrections are known to make nuclear matter more
transparent, we demonstrate that in some instances they lead to an opaqueness.
Numerical calculations are performed for the energies of the HERA-B 
experiment, and the RHIC-LHC colliders.

 \end{abstract}


\maketitle




\section{Introduction}

Large rapidity gap (LRG) events in hadronic collisions at high energies
contain important information about the interaction dynamics. The dominant
contribution to these events comes from colorless gluonic exchange which
is usually associated with diffraction or Pomeron exchange. The
probability to have a colored exchange, but radiate no gluons within a
LRG, attenuates exponentially as function of the rapidity gap, similar to
what is known for secondary Reggeons.

A colorless gluonic exchange may lead to excitation of the colliding 
proton, or the nucleus, or both. We assume the nuclei to be mainly a 
composition of colorless clusters. Therefore, a nucleus may be excited 
either with or without excitation of the clusters. The latter possibility 
is usually called quasielastic or quasidiffractive scattering.

The proton can also be excited via different mechanisms, with or without
excitation of gluonic degrees of freedom. The latter possibility means that the
gluons are not resolved by the interaction, but only the valence quark skeleton
of the proton is excited. This channel of excitation greatly affects the
survival probability for a proton propagating through the nuclear matter. Its
dynamics is very much model dependent. In terms of the Regge phenomenology,
this excitation channel is related to the $\Pom\Pom\Reg$ triple-Regge graph.

The other possibility of proton excitation, due to diffractive gluon
radiation, is usually associated with the triple-Pomeron ($3\Pom$)
mechanism. It has been known since the 70s that the triple-Pomeron
coupling is amazingly small, which means that diffractive gluon
bremsstrahlung is unusually weak. This signals that the gluons in the
proton are located within small spots. Indeed, large mass diffraction
provides a unique access to the process of gluon radiation. The existence
of the $1/M^2$ tail in the effective mass distribution of the
diffractively excited hadron is a clear evidence for radiation of a vector
particle. In order to explain the smallness of the radiation cross section
one should assume that gluonic clouds in the proton are as small as
$0.3\fm$ \cite{kst2,k3p,qm04,shuryak}.

The observed smallness of of the triple-Pomeron diffraction in pp 
collisions also leads to weak gluon shadowing in nuclei \cite{kst2}.
This is indeed confirmed by the latest DGLAP analysis of deep-inelastic 
data in the next-to-leading approximation \cite{florian}.

Thus one should conclude that diffraction, in particular in nuclei, 
provides also sensitive tools for the study of the hadronic structure.

Elastic scattering should be also classified as a LRG process. There is,
however, an important difference between elastic and inelastic
diffraction, in particular, how much they are affected by absorptive
corrections. To see that one can just glance at the data. While the total
and elastic cross sections rise with energy with a rate corresponding to
the Pomeron intercept $\alpha_\Pom(0)\approx 1.1$, the single diffraction
cross section is nearly independent of energy \cite{dino,peter}. This may
be related to the difference in the absorption corrections which look
especially simple in the eikonal approximation and impact parameter
representation,
 \beq
\tilde f_{el}(b) = i\left[1-e^{if_{el}(b)}\right]\ .
\label{5}
 \eeq
 \beq
\tilde f_{sd}(b) = f_{sd}(b)\,e^{if_{el}(b)}\ .
\label{7}
 \eeq
 Here $f_{el}$, $f_{sd}$ and $\tilde f_{el}$, $\tilde f_{sd}$ are the input
and absorption corrected elastic and single diffractive amplitudes
respectively. Expanding the exponentials up to the lowest order correction one
gets a correction factor $[1+\xi if_{el}(b)]$, where $\xi=1/2$ for the elastic
amplitude, Eq.~(\ref{5}), and $\xi=1$ for the diffractive amplitude,
Eq.~(\ref{7}). We conclude that even for the hadronic amplitude the absorption
correction for the diffractive amplitude is twice as big a for the elastic
one. Correspondingly, the result of the absorptive corrections is more
pronounced in diffraction than in elastic scattering. In particular, the
energy rise of the exponent in the absorption factor in (\ref{7}) considerably
slows down the energy dependence of the diffraction cross section. In the case
of diffraction on nuclear targets absorptive corrections produce dramatic
reduction of the cross sections.

Below we demonstrate that the probability of diffractive excitation of the
valence quark skeleton in the proton is also quite small, only $6.5\%$ of
the elastic scattering. However, on the contrary to gluon radiation such a
smallness is due to small overlap of the initial and final wave
functions. Therefore, this probability is sensitive to the proton
structure as well as to the form of the dipole cross section.

In the case of nuclear targets the LRG processes are especially sensitive
to the Gribov inelastic shadowing. Indeed, those corrections affect the
nuclear transparency, i.e. the exponential term in (\ref{5}), (\ref{7}).
For a heavy nucleus this term is tiny, therefore the inelastic corrections
to the elastic amplitude Eq.~(\ref{5}) cannot be large even if the
exponential term changes considerably. At the same time, single
diffraction and other LRG channels are directly affected by inelastic
shadowing and may undergo considerable modification. In what follows, we
perform calculations for different LRG processes within the Glauber model,
as well as within the light-cone dipole approach which incorporates the
inelastic corrections in all orders of multiple interaction. We find that
for quasielastic scattering inelastic shadowing make nuclear matter more
opaque, rather than transparent. This result is at variance with the
conventional wisdom.

There is also a practical reason to perform reliable calculations for the
LRG cross sections. Many experiments at HERA-B, RHIC and future
measurements at LHC have triggering systems which are set up for the central
rapidity region and miss LRG events. For the purpose of normalization one
has to know the part of the inelastic cross section covered by the
trigger, i.e. the total cross section minus the LRG contributions.

The paper is organized as follows. We formulate the light-cone (LC) dipole 
description of soft hadronic reactions in Sect.~\ref{lc}. In particular, 
Sect.~\ref{dc} presents the phenomenological dipole cross section 
fitted to data, and Sect.~\ref{wf} describes the models for the valence 
quark proton wave functions used throughout the paper.

Sect.~\ref{diff} is devoted to the process of single diffractive
excitation in $pp$ collisions. The analysis of data performed in
Sect.~\ref{data} demonstrates a surprisingly small cross section of
diffraction related to excitation of the valence quark skeleton of the
proton.  Attempting to explain this smallness within different models
listed in Sect.~\ref{models} we conclude that only a saturated shape of the
dipole cross section maybe considered realistic. As for the models for
the valence quark distribution in the proton, the truth seems to lie
between the two extremes, the symmetric $3q$ configuration and the
quark-diquark structure with vanishing diquark size.

The triple-Pomeron part of the diffraction related to diffractive gluon
radiation is considered in Sect.~\ref{3-pom}. This contribution is
unambiguously identified in data via its $1/M^2$ tail in the effective
excitation mass distribution. This part of diffraction is well described
by the model, since the quark-gluon LC wave function has been fitted to
this data previously.

Single diffraction, as well as any off-diagonal LRG gap process is subject
to unitarity, or absorptive corrections. These corrections evaluated in
Sect.~\ref{unitarity} substantially slow down the energy dependence of
diffraction.

Unfortunately, the calculation of the double diffractive cross section goes
beyond the employed phenomenology of dipole-proton cross section. It needs
more refined information about the dipole-dipole cross section.  
Therefore, we make a simple estimate of this cross section in
Sect.\ref{dd}, based on Regge factorization.

In Sect.~\ref{glauber} we switch to proton-nucleus collisions and present the
Glauber model approach to LRG processes. Unfortunately most of LRG channels
cannot be accessed in this approximation which neglects all off-diagonal
diffractive reactions. Therefore, one should include the Gribov inelastic
shadowing corrections which are introduced in Sect.~\ref{ct} within the
eigenstate formalism. It turns out that at high energies the interaction
eigenstates are the color dipoles, thus the Gribov corrections lead to the
color transparency effect \cite{zkl}.

In Sect.~\ref{coherent} cross sections of coherent processes in which the
nucleus remains intact are calculated. These include elastic scattering and
diffractive excitation of the projectile proton, as well as the total cross
section related via unitarity to the elastic amplitude. We employ different
combinations of the two models for the proton wave function and two models
for the dipole cross section. The results are presented in
Table~\ref{hera1} for the energy of HERA-B, $\sqrt{s}=41.6\GeV$, in
Table~\ref{rhic} for RHIC, $\sqrt{s}=200\GeV$, and in Table~\ref{lhc} for
LHC, $\sqrt{s}=5.5\TeV$

The reactions leading to nuclear break-up, quasielastic scattering and
diffractive excitation of bound nucleons are considered in
Sect.~\ref{incoherent}. One can calculate the cross section making use of
completeness.

Besides the excitation of the proton valence quark skeleton, diffractive
interaction can shake gluons off the proton. In terms of Regge phenomenology
this process is related to the triple Pomeron part of diffraction. On the
other hand, it causes a reduction of gluon density in the proton, an effect
usually called gluon shadowing. These phenomena are considered in
Sect.~\ref{gluons}.

\section{Light-cone dipole representation}\label{lc}

\subsection{The dipole cross section}\label{xsect}\label{dc}

The cross section of interaction of a $\bar qq$ dipole with a nucleon,
introduced in \cite{zkl}, is usually assumed to be flavor independent, and to
depend only on the $\bar qq$ transverse separation and energy. This
independence of quark flavor has been proven only perturbatively, but good
agreement between data and a parameter free calculation \cite{ihkt} of the
$J/\Psi$ photoproduction cross section shows that this is a good
approximation.

The dipole cross section calculated perturbatively \cite{zkl} within the
two-gluon model for the Pomeron \cite{low,nussinov} depends only on the
transverse dipole size $r_T$ and is independent of energy and the relative
fractions of the momentum carried by the quark and antiquark. However, higher
order corrections bring forth a dependence on energy (or Bjorken
$x$-dependence in DIS) and on the fractions $\alpha$ and $1-\alpha$ of the
dipole light-cone momentum carried by the $q$ and $\bar q$. This energy
dependence is rather mild (especially in soft processes under consideration)
and may lead to sizeable effects only for large variations of the collisions
energy. Since the variation of $\alpha$ is usually not very large, its
dependence can be neglected. This approximation may be questionable in DIS
where the end-point region of the $\alpha$ distribution is enhanced by the
increasing cross section \cite{kp}. However, this is not the case in soft
reactions where the $\alpha$-independence should be a good approximation.

In what follows we will test two popular models for the dipole cross 
section: \\ \\
{\it\bf (i) Simple extrapolation of the 
small-\boldmath$r_T$ 
behavior,}
\beq
\sigma_{\bar qq}(r_T,s) = C(s)\,r_T^2\ ;
\label{170}
 \eeq
 {\it\bf (ii) The saturated cross section rises as \boldmath$r_T^2$
at small \boldmath$r_T^2$, but levels off at large \boldmath$r_T^2$,}
 \beq
\sigma_{\bar qq}(r_T,s)=\sigma_0(s)\,\left[
1-{\rm exp}\left(-\frac{r_T^2}
{R_0^2(s)}\right)\right]\ ,
\label{180}
 \eeq
 where $R_0(s)=0.88\,fm\,(s_0/s)^{0.14}$ and $s_0=1000\,GeV^2$
\cite{kst2}. The energy dependent factor $\sigma_0(s)$ is defined as,
 \beq
\sigma_0(s)=\sigma^{\pi p}_{tot}(s)\,
\left(1 + \frac{3\,R^2_0(s)}{8\,\la r^2_{ch}\ra_{\pi}}
\right)\ ,
\label{190}
 \eeq
 where $\la r^2_{ch}\ra_{\pi}=0.44\pm 0.01\,fm^2$ \cite{pion} is the mean
square of the pion charge radius.  

This dipole cross section is normalized to reproduce the pion-proton total
cross section, $\la\sigma_{\bar qq}\ra_\pi=\sigma_{tot}^{\pi p}(s)$. The
saturated shape of the dipole cross section is inspired by the popular
parametrization given in Ref.~\cite{gbw}, which is fitted to the low-$x$ and
high $Q^2$ data for $F^p_2(x,Q^2)$ from HERA. However, that should not be used
for our purpose, since is unable to provide the correct energy dependence of
hadronic cross sections. Namely, the pion-proton cross section cannot exceed
$23\mb$ \footnote{According to \cite{sgbk} this dipole cross section
reproduces well the energy dependence of the photoabsorption cross section
$\sigma^{\gamma p}_{tot}(s)$. This happens only due to the singularity in the
light-cone wave function of the photon at small $r_T$, which is a specific
property of this wave function and is not applicable to hadrons.}. Besides,
Bjorken $x$ is not a proper variable for soft reactions, since at small $Q^2$
the value of $x$ is large even at low energies. The $s$-dependent dipole cross
section Eq.~(\ref{180}) was fitted \cite{kst2} to data for hadronic cross
sections, real photoproduction and low-$Q^2$ HERA data for the proton
structure function.  The cross section (\ref{170}) averaged with the pion wave
function squared (see below) automatically reproduces the pion-proton cross
section.

In the case of a proton beam one needs a cross section for a three-quark dipole,
$\sigma_{3q}(\vec r_1,\vec r_2,\vec r_3)$, where $\vec r_i$ are the transverse
quark separation with a condition $\vec r_1+\vec r_2+\vec r_3=0$. In order to
avoid the introduction of a new unknown phenomenological quantity, we 
express the
three-body dipole cross section via the conventional dipole cross section $\sq$
\cite{mine},
 \beq
\sigma_{3q}(\vec r_1,\vec r_2,\vec r_3) =
{1\over2}\,\Bigl[\sigma_{\bar qq}(r_1)+
\sigma_{\bar qq}(r_2)+
\sigma_{\bar qq}(r_3)\Bigr]\ .
\label{195}
 \eeq
 This form satisfies the limiting conditions, namely, turns into
$\sigma_{\bar qq}(r)$ if one of three separations is zero. Since all these
cross sections involve nonperturbative effects, this relation hardly can
be proven, but should be treated as a plausible assumption.

\subsection{The light-cone wave function of the proton}\label{wf}

Since the dipole cross section is assumed to be independent of the sharing
of the light-cone momentum among the quarks, the wave function squared of
the valence Fock component of the proton, $\left|\Phi(\vec
r_i,\alpha_j\right|^2$ should be integrated over fractions $\alpha_i$. The
result depends only on transverse separations $\vec r_i$. The form of the
nonperturbative valence quark distribution is unknown, therefore for the
sake of simplicity we assume the Gaussian form,
 \beqn
&&\left|\Psi_N(\vec r_1,\vec r_2,\vec r_3)\right|^2 
\nonumber\\ &=&
\int\limits_0^1 \prod\limits_{i=1}^3 d\alpha_i\,
\left|\Phi(\vec r_i,\alpha_j)\right|^2\,
\delta\left(1-\sum\limits_{j=1}^3 \alpha_j\right)
\nonumber\\ &=&
\frac{2+r_p^2/R_p^2}{(\pi\,r_p\,R_p)^2}
\exp\left(-\frac{r_1^2}{r_p^2}-\frac{r_2^2+r_3^2}{R_p^2}\right)\,
\nonumber\\ &\times&
\delta(\vec r_1+\vec r_2+\vec r_3)\ ,
\label{200}
 \eeqn
 where $\vec r_i$ are the interquark transverse distances. The two scales
$r_p$ and $R_p$ characterizing the mean transverse size of a diquark and
the mean distances to the third quark. In what follows we consider two
extreme possibilities:

{\bf (i) The quark-diquark structure of the proton.} There are evidences
from data that the dominant configuration of valence quarks in the proton
contains a small isoscalar $ud$ diquark of a size $r_p \sim 0.2-0.4\fm$
\cite{sb,diquark,kz}, i.e. $r_p^2\ll R_p^2$. Neglecting the diquark radius
we arrive at a meson-type color dipole structure,
 \beq
\left|\Psi_N(\vec r_1,\vec r_2,\vec r_3)\right|^2 = 
\frac{2}{\pi\,R_p^2}
\exp\left(-\frac{2\,r_T^2}{R_p^2}\right)\ ,
\label{210}
 \eeq
 where $\vec r_T=\vec r_2=\vec r_3$, and $R_p$ is related to 
the mean charge radius squared of the proton as $R_p^2={16\over 3}\la 
r^2_{ch}\ra_p$.

{\bf (ii) The symmetric proton structure.} Another extreme would be to say
that the forces binding the valence quarks are of an iso-scalar nature,
therefore the quark distribution is symmetric, i.e. $r_p=R_p$ in
(\ref{200}). In this case the mean interquark separation squared is
$\la \vec r_i^{\,2}\ra={2\over3}R_p^2 = 2\la r_{ch}^{2}\ra_p$.

Thus, we consider two possibilities for the shape of the dipole cross 
section and two choices for the proton wave function. In what follows we 
will test these models comparing to data.

\section{Single diffraction \boldmath$pp\to Xp$}\label{diff}

Now we have a choice for the key ingredients of the dipole approach
\cite{zkl}, which are the universal dipole cross section and the
light-cone wave function of the incoming hadron. First of all, we should
make sure that each model correctly reproduces the total $pp$ cross
section, which is the main entry for the calculation of nuclear effects.  
A further important and rigorous condition for a realistic model would be
the magnitude of the single diffractive cross section.

\subsection{Experimental data}\label{data}

The cross section of proton excitation, $pp\to pX$, integrated over large
Feynman $x_F>0.95$ is about $3.5\mb$ at $E_{lab}=920\GeV$ and nearly
saturates at the value of $4\mb$ at higher energies \cite{dino,peter}.
This cross section, however, includes two parts: (i) excitation of the
quark skeleton of the hadron without gluon radiation; (ii) Excitations
including emission of gluons. The former, in terms of the Regge
phenomenology, corresponds to the triple Regge graph $\Pom\Pom\Reg$ and
behaves like $1/M^3$ at large excitation masses. The latter corresponds to
the triple-Pomeron graph $3\Pom$ which provides the $1/M^2$ tail at
large masses.

In terms of the dipole description the lowest Fock component at 
which we are concentrating in this section corresponds 
to the first part, $\Pom\Pom\Reg$. The higher Fock components containing 
gluons will be considered in Sect.~\ref{gluons}.

At high energies and large $x_F\to 1$ one can write the diffractive cross 
section in the triple-Regge form \cite{3R},
 \beqn
&& \frac{d\sigma_{sd}^{pp}}
{dx_F\,dp_T^2} = 
\sqrt{\frac{s_1}{s}}\ 
\frac{G_{\Pom\Pom\Reg}(0)}
{(1-x_F)^{3/2}}\ 
e^{-B_{3\Reg}p_T^2}
\nonumber\\ &+&
\frac{G_{3\Pom}(0)}
{(1-x_F)}\ 
e^{-B^{pp}_{3\Pom}p_T^2} +
G_{\Reg\Reg\Pom}(0)\ 
e^{-B_{\Reg\Reg\Pom}p_T^2}\ .
\label{140}
 \eeqn
 Here $s_1=1\GeV^2$,
\beqn
B_{\Pom\Pom i} &=&
R^2_{\Pom\Pom i} -
2\alpha^\prime_\Pom\,\ln(1-x_F)\nonumber\\
B_{\Reg\Reg\Pom} &=&
R^2_{\Reg\Reg\Pom} -
2\alpha^\prime_\Reg\,\ln(1-x_F)\ ,
\label{150}
 \eeqn
 $i=\Pom,\Reg$;  $\alpha^\prime_\Pom\approx 0.25\GeV^{-2}$ and
$\alpha^\prime_\Reg\approx 0.9\GeV^{-2}$ are the slopes of the Pomeron
and secondary Reggeon trajectories. 
Note that we keep in (\ref{150}) only the triple-Regge terms which are
either singular in $(1-x_F)$, or do not vanish as powers of energy.

Since the data show no substantial rise of the diffractive cross section with
energy \cite{dino,peter}, which is apparently caused by strong absorptive
corrections, we incorporate this fact fixing the effective Pomeron
intercept at $\alpha_\Pom(0)=1$. This also allows us to use the results of
the comprehensive triple-Regge analysis of data performed in
Ref.~\cite{3R}, in which $\alpha_\Pom(0)=1$ was used. They found the
following values for the parameters: $G_{3\Pom}(0)=
G_{\Pom\Pom\Reg}(0)=3.2\mb/\GeV^2$;  
$G_{\Reg\Reg\Pom}(0)=13.2\mb/\GeV^2$; $R^2_{3\Pom}=4.2\GeV^{-2}$;
$R^2_{\Pom\Pom\Reg}=1.7\GeV^{-2}$; $R^2_{\Reg\Reg\Pom}=0\GeV^{-2}$.

We are now in a position to evaluate the diffractive cross section integrating
the distribution Eq.~(\ref{140}) over $p_T^2$ and $x_F$, from $x_{min}$ up to
$x_{max}$. We fix $x_{min}=0.95$ as is usually done to separate diffractive
from nondiffractive contributions, and $x_{max}=1-M_0^2/s$ with
$M_0^2=(m_N+m_\pi)^2\GeV^2$.  Of course such small masses do not satisfy the
condition of the triple-Regge dynamics, but we appeal to the duality between
$s$-channel resonances and $t$-channel Reggeons which is known to work well if
one averages over the resonances. Then we get at the energy of HERA-B,
$\sigma_{sd}^{pp}=3.27$ which agrees well with the value
$\sigma_{sd}^{pp}=3.5\mb$ suggested by data. We expect
$\sigma_{sd}^{pp}\approx 4\mb$ at RHIC and LHC energies.

As was already mentioned, in this section we are only interested in the
part of diffraction related to excitation of the valence quark skeleton
without gluon radiation. This part corresponds to the first term,
$\Pom\Pom\Reg$, in Eq.~(\ref{140}). Integrating it over $p_T^2$ and $x_F$
we get $\sigma^{pp}_{sd}(\Pom\Pom\Reg) = 1.13\mb$, i.e. about the third of
the total single diffractive cross section.

For further model tests we need to know the forward diffraction which
turns out to be quite small with respect to the forward elastic cross
section,
 \beq
R_{sd}=\left.\frac{d\sigma_{sd}/dp_T^2}
{d\sigma_{el}/dp_T^2}\right|_{p_T=0} =
\frac{5.5\mb/\GeV^2}{84.5\mb/\GeV^2}=0.065\ ,
\label{160}
 \eeq
 at the energy of HERA-B.

 It is worth reminding that we rely on the hypothesis of duality and
should not expect this result to be very accurate. It just shows the scale
of the effect, namely, that diffractive excitation of the projectile
valence quark system is an order of magnitude weaker than the elastic
channel.

 In hadronic representation, in particular within the one pion exchange
model, one can explain rather well such a weak diffractive excitation of
the valence Fock component of the proton \cite{3R}. It is a challenge,
however, to reproduce this result within the QCD based dipole model.

\subsection{Models}\label{models}

Since we have models for the dipole cross section and the proton wave
function, it is therefore possible to calculate the total and forward single
diffractive cross sections \cite{zkl,kst2},
 \beq
\sigma^{pp}_{tot} =
\left\la\sq(r_i)\right\ra\ ;
\label{164}
 \eeq

 \beqn
&& \left.\int dM_X^2\,
\frac{d\sigma_{sd}(pp\to pX)}{dM_X^2\,dp_T^2}\right|_{p_T=0}
\nonumber\\ &=& 
\frac{1}{16\pi}\left\{
\Bigl\la[\sq(r_i)]^2\Bigr\ra -\Bigl\la\sq(r_i)\Bigr\ra^2\right\}\ .
\label{165}
 \eeqn
 Here the averaging weight is the proton wave function squared,
 \beq
\la ...\ra = \int \prod\limits_{i=1}^3 d^2r_i\,
|\Psi(r_j)|^2\,(...)\ .
\label{166}
 \eeq
 We are going to test the models mentioned before regarding their ability
to explain the strong suppression of the diffractive relative to the
elastic channels, $R_{sd}\sim 0.1$ in Eq.~(\ref{160}).

\subsubsection{Model I: quark-diquark proton and \boldmath$\sq\propto 
r_T^2$}\label{var1}

The simple quadratic $r_T$ dependence of the dipole is quite popular in the
literature and is frequently used for calculations of nuclear effects (e.g.
see \cite{km,mine,kkt}). Since the energy dependent factor is unknown, it is
usually fitted to reproduce the total cross section. In this case one has to
explain the relative values of total cross sections for different hadronic
species, in particular, the ratio $\sigma^{pp}_{tot}/\sigma^{\pi
p}_{tot}\approx 1.6$. Indeed, the ratio of charged radiuses squared agrees
with this value within the diversity of the results of different measurements
\cite{r-ch}.

However, the cross section of single diffraction turns out to be
dramatically overestimated. Indeed, the diffraction to elastic ratio
defined in (\ref{160}) reads,
 \beq
R_{sd}= \frac{\la r_T^4\ra}{\la r_T^2\ra^2} - 1 = 1\ .
\label{220}
 \eeq This is more than one order of magnitude larger than the
experimental value Eq.~\ref{160}.

\subsubsection{Model II: symmetric proton and \boldmath$\sq\propto 
r_T^2$}\label{var2}

In this case we again treat the unknown factor in the dipole cross 
section as a free parameter and adjust it to the total cross section,
 \beq
\left\la \sigma_{3q}(\vec r_i)\right\ra 
= \sigma_{tot}^{pp}\ ;
\label{230}
 \eeq
 Then we can calculate the mean value of the dipole cross section squared,
 \beq
\left\la \sigma^2_{3q}(\vec r_i)\right\ra =
{3\over2}\,\left(\sigma_{tot}^{pp}\right)^2\ ,
\label{240}
 \eeq
 and eventually the diffractive cross section according to (\ref{165}).
Then we arrive at the ratio,
 \beq
R_{sd}={1\over2}\ ,
\label{250}
 \eeq
 which is closer to the experimental value Eq.~(\ref{160}), but still 
quite overestimates the data.

It is not a surprise that both above models based on the dipole cross
section Eq.~(\ref{170}) steeply rising with quark separation grossly
overestimate the diffractive cross section, since the main contribution
comes from large distances. Therefore, both models I and II are not
realistic and should not be used for comparison with data.

Actually, Eq.~(\ref{165}) shows that diffraction is given by the
dispersion of the dipole cross section dependence on the transverse $\bar
qq$ separation.  Apparently, $\sigma_{\bar qq}(r_T)\propto r_T^2$ used so
far varies with $r_T$ too steeply and the dispersion is too large. One
needs a flatter dependence in order to suppress diffraction.

\subsubsection{Model III: quark-diquark proton and saturated cross
section}\label{var3}

There is no freedom in this case, the dipole cross section Eq.~(\ref{180})
if fixed to reproduce the pion-proton cross section. The calculated value
of proton-proton total cross section is rather close to the experimental
value, but somewhat smaller. This is not a surprise, since this model
employs the approximation of a point-like diquark, which apparently leads
to an underestimation of the $pp$ cross section.  It is risky, however, to
rely on such an approximation for calculating nuclear effects. Even a
small deviation of $\sigma^{pp}_{tot}$ from the experimental value will
cause nuclear effects which may be misinterpreted as inelastic shadowing.  
To make sure that this cross section is exactly reproduced, we redefine
the function $\sigma_0(s)$ in (\ref{190}),
 \beq
\sigma^{III}_0(s)=\sigma^{pp}_{tot}(s)
\left(1+{1\over\delta}\right)\ ,
\label{255}
 \eeq
 where the parameter $\delta=8\la r_{ch}^2\ra_p/3R_0^2(s)$, and we use the
mean proton charge radius squared $\la r_{ch}^2\ra_p=0.8\fm^2$ \cite{r-ch}.

Using the proton wave function Eq.~(\ref{210}) and the saturated cross 
section Eq.~(\ref{180}) we get the single diffractive cross section in the 
form,
 \beq
\left.\frac{d\sigma_{sd}^{pp}}{dp_T^2}\right|_{p_T=0}=
\frac{\delta^2}{(1+\delta)^2(1+2\delta)}\ 
\frac{\sigma_0^2(s)}{16\pi}\ .
\label{260}
 \eeq

Dividing by the elastic cross section we get,
 \beq
R_{sd} = \frac{1}{1+2\delta}\ .
\label{280}
 \eeq
 This ratio is rather small at the energy of HERA-B, $R_{sd}= 0.13$, which
is compatible with the data. The fraction of single diffraction decreases
with energy down to $R_{sd}=0.06$ at the energy of RHIC
($\sqrt{s}=200\,$GeV), and $R_{sd}=0.01$ at the energy of LHC
($\sqrt{s}=5.5\,$TeV).

One may wonder why the saturated cross section leads to such a weak
diffraction and why decreases with energy? This is easy to interpret.
Indeed, if the cross section were completely flat, i.e. $\sq(r_T)=const$,
no diffraction would be possible because of orthogonality of the initial
and final valence quark wave functions. Only the drop of $\sq(r_T)$ at
$r_T\lsim R_0(s)$ makes diffraction possible. However, $R_0(s)$ decreases
with energy, therefore the shape of the dipole cross section is getting
flatter and diffraction vanishes. As far as diffraction gets its main
contribution from small $r_T\lsim R_0$, note that it is less probable to
find three quarks with small separations, than a two-quark system.
Therefore, one should expect less diffraction for a symmetric three-quark
wave function of the proton as is demonstrated in the next section.

\subsubsection{Model IV: symmetric proton and saturated 
cross section}\label{var4}

Using the wave function Eq.~(\ref{200}) with $r_p=R_p$ and the cross
section (\ref{180}) we get the following forward elastic cross section
and diffraction to elastic ratio,
 \beqn
\left.\frac{d\sigma^{pp}_{el}}
{dp_T^2}\right|_{p_T=0} &=&
\frac{\gamma^2}{(1+{2\over3}\gamma)^2}\ 
\frac{\sigma_0^2(s)}{16\pi}\ ;
\label{265}\\
R_{sd}&=&
\frac{(1+{1\over3}\,\gamma)^2-{1\over2}}
{(1+{1\over3}\,\gamma)(1+{4\over3}\,\gamma)
(1+\gamma)}\ ,
\label{290}
 \eeqn
 The parameter $\gamma$ is related to previously introduced $\delta$,
 \beq
\gamma = 3\,\frac{\la r_{ch}^2\ra_p}{R_0^2(s)}
= {9\over8}\,\delta\ .
\label{295}
 \eeq
 Then, for the energy of HERA-B we arrive at a very small fraction of the
single diffractive cross section $R_{sd}=0.07$ which is in excellent
agreement with the experimental result Eq.~(\ref{160}). We expect a
substantial reduction of this fraction at higher energies, $R_{sd}\approx
0.03$ at the energy of RHIC, and $R_{sd}\approx 0.0045$ at the energy of
LHC.

Note that the the parameter $\gamma$ in Eqs.~(\ref{265})-(\ref{290}) can 
be defined differently,
 \beq
\sigma^{pp}_{tot} = \left\la \sigma_{3q}\right\ra =
\sigma_0(s)\,\frac{\gamma}{1+{2\over3}\,\gamma}\ ,
\label{300}
 \eeq
 where $\sigma_0(s)$ and $\sigma_{3q}(\vec r_i)$ are defined in
(\ref{190}) and (\ref{195}) respectively. 

Both definitions (\ref{295}) and (\ref{300}) nicely agree at the energies
of fixed target experiments. More problematic is to apply Eq.~(\ref{300})
at high energies of colliders, RHIC and LHC. No data for $\sigma^{\pi
p}_{tot}$ are available at energies above $\sqrt{s}=30\GeV$. The usual
extrapolation with a universal energy dependence as in $pp$
collisions is just an educated guess not supported by any dynamic theory.
Moreover, one should expect a steeper rise of $\sigma^{\pi p}_{tot}$ than
$\sigma^{pp}_{tot}$. Indeed, the rising part of the cross section related
to gluon radiation is nearly independent of hadronic size, while the
constant part of the cross section related to dipole-dipole collision
followed by no gluon radiation apparently depends on the dipole sizes and
is smaller for $\pi p$, than for $pp$. Thus, pion-proton cross section
rises steeper with energy, in accordance with the general trend of steeper
energy dependence for smaller dipoles discovered at HERA.

One can combine Eqs.~(\ref{295}) and (\ref{300}) in order to find the
unknown energy dependence of $\sigma^{\pi p}_{tot}(s)$. This is a more
reliable procedure, since the dipole cross section, in particular the
parameter $R_0(s)$, is fitted to data at energies much higher than those
where data for $\sigma^{\pi p}_{tot}(s)$ are available. Then the pion to
proton ratio of the cross sections reads,
  \beq
R_{\pi/p}= \frac{\sigma^{\pi p}_{tot}(s)}
{\sigma^{pp}_{tot}(s)} =
\frac{{2\over3}+{1\over3}\,
\frac{R_0^2(s)}{\la r_{ch}^2\ra_p}}
{1+{3\over8}\,
\frac{R_0^2(s)}{\la r_{ch}^2\ra_{\pi}}}\ .
\label{301}
 \eeq
 This ratio slowly rises with energy. At the energy of HERA-B
$R_{\pi/p}=0.6$ in excellent agreement with data, but it becomes $10\%$
larger, $R_{\pi/p}=0.66$, at the energy of LHC. Eventually, at very high
energy when unitarity will be saturated, all the cross sections reach the
universal Froissart limit corresponding to an expanded black disk.
Correspondingly, $R_{\pi/p}(s)\to 1$.

\subsection{Diffractive gluon radiation}\label{3-pom}

Besides excitation of the valence quark skeleton of the proton, a valence
quark itself can be excited followed by gluon radiation. In terms of
Regge phenomenology this process corresponds to the triple-Pomeron
contribution to the diffraction cross section. This is the second term in
Eq.~(\ref{140}).

It has been known since the 70s that the triple-Pomeron coupling is quite
small. To appreciate this statement one can express diffraction in terms of
the Pomeron-proton total cross section which should be expected to be twice as
large as a meson-proton one. Indeed, the Pomeron is a gluonic system,
therefore one should have an extra Casimir factor $9/4$ compared to a quark
dipole. Thus, one may expect $\sigma_{tot}^{\Pom p}\approx 50\mb$. This
expectation is in dramatic contradiction with data \cite{dino} which show only
$\sigma_{tot}^{\Pom p} = 2\mb$ at large excitation masses. Apparently, it is
much more difficult to shake gluons off valence quarks, compared to pQCD
expectations. The way out of this puzzle is to suggest that gluons in the
proton are located within small spots which are hardly resolved by soft
interactions. The mean transverse size of these spots was fitted to single
diffraction data with large effective masses and found to be $\la r_T\ra\sim
r_0=0.3\fm$. Such gluons have much more intensive Fermi motion than massless
perturbative ones, and they are less sensitive to an external kick, i.e. gluon
radiation is suppressed.

Such a picture is quite successful in explaining the energy dependence of
the total hadronic cross sections and elastic hadronic slopes \cite{k3p}.
It also correctly predicts the reduced value of $\alpha_{\Pom}^\prime$,
the slope of the Pomeron trajectory for the process of elastic
photoproduction of $J/\Psi$ \cite{qm04}.  Similar statements about gluonic
structure of the proton has been done in \cite{shuryak} recently.

The light-cone wave function of the quark-gluon Fock component of a quark was 
calculated in \cite{kst2} within a model describing the nonperturbative 
interaction of gluons via a phenomenological light-cone potential taken in 
an oscillatory form. The result reads,
 \beq 
\Psi_{qG}(\vec r_T)= -\frac{2i}{\pi}\,
\sqrt{\frac{\alpha_s}{3}}\
\frac{\vec r_T\cdot\vec e^{\,*}}{r_T^2}\, 
{\rm exp}\left(-{r_T^2\over2r_0^2}\right)\ . 
\label{302}
 \eeq
 Here we assume that the gluon, which is a vector particle and possesses a
$1/x$ distribution, is carrying a negligible fraction $\la x\ra\ll1$ of the
quark momentum. Of course the concrete shape of the light-cone potential is
not crucial. What is only important is the smallness of the mean quark-gluon
separation. Notice that Eq.~(\ref{302}) can be viewed as a source of small
gluonic spots in the proton \cite{qm04}.

Since the light-cone quark-gluon distribution function is known, one
can calculate the cross section of diffractive gluon radiation by a
high-energy quark interaction with a nucleon \cite{kst2},
 \beqn   
&&\frac{d\sigma(qN\to qGN)}{dx_F\,dp_T^2}
\biggr|_{p_T=0} = \frac{1}{16\pi(1-x_F)}
\nonumber\\ &\times&
\int d^2r_T\,\biggl|\Psi_{qG}(\vec r_T)\biggr|^2\,
\widetilde\sigma^2(r_T,s)\ .
\label{303}
 \eeqn
 Here the cross section $\widetilde\sigma(r_T,s)$ is not just a
cross section of interaction of a $qG$ dipole. This dipole is not even
colorless. As usual, diffractive excitation is possible due to a
difference between elastic amplitudes for different Fock states, in this
case the bare quark $|q\ra_0$ and the $|qG\ra$ pair. Since they have the
same color, the difference emerges from the color-dipole moment of the
$q-G$ pair. It is shown in \cite{kst2} (see in particular Appendix~A.2)  
that $\widetilde\sigma(r_T,s)={9\over8}\,\sq(r_T,s)$.

The next step is to integrate over $p_T$ the cross section of diffractive
gluon radiation provided that the forward one, Eq.~(\ref{303}), is known.
In terms of the Regge phenomenology diffractive radiation corresponds to
the triple-Pomeron term in the cross section of single diffraction. Data
agree with a Gaussian $p_T$-dependence of the triple-Pomeron term with the
slope \cite{3R},
 \beq
B^{pp}_{3\Pom}(x_F) = B_{3\Pom}^0+2\,\alpha_{\Pom}^\prime\,
\ln\left(\frac{1}{1-x_F}\right)\ ,
\label{304}
 \eeq
 where $B_{3\Pom}^0=4.2\GeV^{-2}$, and $\alpha_{\Pom}^\prime = 
0.25\GeV^{-2}$.

Now, we are in a position to evaluate the effective triple-Pomeron part of
the single diffraction cross section for $pp$ collisions employing the the
wave function Eq.~(\ref{302}) and the saturated cross section
Eq.~(\ref{180}),
 \beqn
&& \left[\frac{d\sigma(pp\to pX)}{dx_F\,dp_T^2}\right]_{3\Pom}
= \frac{81\,\alpha_s\,\sigma_0^2(s)}{(16\pi)^2(1-x_F)}
\nonumber\\ &\times&
\ln\left[1+\frac{\epsilon^2(s)}{1+2\epsilon(s)}\right]\,
\exp\left[-p_T^2\,B^{pp}_{3\Pom}(x_F)\right]\ ,
\label{305}
 \eeqn
 where $\epsilon(s)=r^2_0/R^2_0(s)$. In the energy range under discussion
$\epsilon(s)$ is rather small, then the single diffractive cross section
Eq.~(\ref{305}) is proportional to $r_0^4$. This is why this process is
quite sensitive to the value of $r_0$ and provides an efficient way to
determine the size of gluonic spots in the proton, $r_0\approx 0.3\fm$
\cite{kst2}.

Notice that the interference between the diffractive amplitudes of gluon
radiation by different quarks in the proton should not be appreciable, 
since $r_0$ is small compared to the proton radius. Explicit calculations 
performed in \cite{kst2} confirm this.

\subsection{Unitarity corrections}\label{unitarity}

Any large rapidity gap process is subject to unitarity or absorptive
corrections which may be substantial. Indeed, initial/final state
interactions tend to fill the rapidity gap by producing particles, and one
may treat such corrections as a survival probability of the rapidity gap.
Such corrections become especially large and may completely terminate the
gap in the vicinity of the unitarity limit, which is also called black
disk regime.

Since the phenomenological dipole cross section $\sq(s)$ is fitted to
data, we assume it to include by default all the unitarity corrections.
The same is true for the off-diagonal amplitude of diffractive excitation
of the valence quark system without gluon radiation, since it is a linear
combination of elastic dipole amplitudes. Thus, our calculations for the
$\Pom\Pom\Reg$ term in (\ref{140}) do not need any unitarity corrections.

The situation is different for the $\Pom\Pom\Pom$ term, or gluon
diffractive radiation. Although the amplitude of diffractive excitation of
a quark, $qN\to qGN$, does include all the absorptive corrections
contained in the phenomenological dipole cross section, the presence of
other projectile valence quarks, the spectators, should not be ignored.
Indeed, any inelastic interactions of the large-size three quark dipole in
the proton, will cause particle production which will fill the rapidity
gap. Thus, one may expect large absorptive corrections to the cross
section of diffractive gluon radiation.

Data for elastic $pp$ scattering show that the partial amplitude
$f^{pp}_{el}(b,s)$ hardly varies with energy at small impact parameters
$b\to 0$, while rises as function of energy at large $b$
\cite{amaldi,kp1,k3p}. This is usually interpreted as a manifestation of
saturation of the unitarity limit, $\Im f^{pp}_{el}\leq 1$. Indeed, this
condition imposes a tight restriction at small $b$, where $\Im
f^{pp}_{el}\approx 1$, leaving almost no room for further rise.
Correspondingly, the amplitude of any off-diagonal process including
single triple-Pomeron diffraction acquires a suppression factor
 \beq
f^{pp}_{sd}(b,s)\Rightarrow
f^{pp}_{sd}(b,s)\,\left[1-\Im f^{pp}_{el}(b,s)\right]\ ,
\label{306}
 \eeq
 due to unitarity or absorptive corrections. This factor related to the
survival probability of LRG is known to decrease with energy \cite{glm}.
Interply of the rising and falling energy dependences of the two factors in
(\ref{306}) may explain the observed flat behavior of the single diffractive
cross section \cite{dino,peter}.

 Since $\Im f^{pp}_{el}(b,s)$ is known directly from data, it would be
straightforward just to fit the data with any proper parametrization and
use the result in Eq.~(\ref{306}). Alternatively, one can use any model
which provides a good description for $\Im f^{pp}_{el}(b,s)$. It is
demonstrated in \cite{k3p} that even the simple model treating the Pomeron
as a Regge pole with no unitarity corrections describes reasonably well
not only the total hadronic cross sections, but even the data
for $f^{pp}_{el}(b,s)$. In this model,
 \beq
\Im f^{pp}_{el}(b,s)=
\frac{\sigma^{pp}_{tot}(s)}
{4\pi B^{pp}_{el}(s)}\ 
\exp\left[-\frac{b^2}{2
B^{pp}_{el}(s)}\right]\ .
\label{307}
 \eeq
 Here and for further applications we use the parametrization from
\cite{pdg}, $\sigma^{pp}_{tot}(s) = 18.76\mb\times(s/M_0^2)^{0.093} +
\sigma^{pp}_R(s)$, where $M_0=1\GeV$, and the Reggeon part of the cross
section $\sigma^{pp}_R(s)$ is small at high energies and can be found in
\cite{pdg}. The elastic slope $B^{pp}_{el}(s) = B^0_{el} +
2\,\alpha^\prime_{\Pom}\, \ln(s/M_0^2)$ with $B^{0}_{el}=8.9\GeV^{-2}$ and
$\alpha^\prime_{\Pom}=0.25\GeV^{-2}$.  Note that the elastic amplitude at
small impact parameters, i.e. the pre-exponential factor in (\ref{307}),
hardly changes with energy imitating saturation of unitarity. This fact
has been known back in the 70s as a geometrical scaling. It is
demonstrated in \cite{k3p} (see Fig.~9) that not only at $b=0$, but in the
whole range of impact parameters the model Eq.~(\ref{307}) reasonably well
describes the energy dependence of the partial amplitude
$f^{pp}_{el}(b,s)$.

Using Fourier transformed Eq.~(\ref{307}) we arrive at the following cross
section for single diffraction integrated over momentum transfer,
 \beqn
&&\left[\frac{d\sigma(pp\to pX)}{dx_F}\right]_{3\Pom}
= \frac{81\,\alpha_s\,\sigma_0^2(s)}
{(16\pi)^2(1-x_F)\,B^{pp}_{3\Pom}(x_F)}\ 
\nonumber\\ &\times&
\ln\left[1+\frac{\epsilon^2(s)}{1+2\epsilon(s)}\right]\,
\left\{1-{1\over{\pi}}\,\frac{\sigma^{pp}_{tot}(s)} 
{B^{pp}_{3\Pom}(x_F)+2B^{pp}_{el}(s)}\right. 
\nonumber\\ &+& \left.
\frac{1}{(4\pi)^2}\,
\frac{\left[\sigma^{pp}_{tot}(s)\right]^2}
{B^{pp}_{el}(s) 
\left[B^{pp}_{3\Pom}(x_F)+B^{pp}_{el}(s)\right]}\right\}\ .
\label{308}
 \eeqn

 At the scale corresponding to the mean transverse momentum of gluons,
$\alpha_s(1/r^2_0)\approx 0.4$ \cite{k3p}, (\ref{308})  agrees with
data reasonably well. Nonetheless, we think that one should perform more
elaborated calculations when $\Im f_{el}(s)$ is close to the unitarity
limit. Indeed, the absorptive correction factor in (\ref{306}) is so
small in this case that any little variations of $\Im f_{el}(s)$ may lad
to dramatic effects.  Such a fine tuning goes beyond the scope of the
paper, but we plan to work more on this elsewhere.

\subsection{Double diffraction \boldmath$pp\to XY$}\label{dd}
 
If the Pomeron were a true Regge pole, one would expect Regge
factorization relating the forward single and double diffractive cross
sections,
 \beq
\left(\frac{d\sigma^{pp}_{dd}} 
{dt}\right)_{t=0}
\left(\frac{d\sigma^{pp}_{el}}
{dt}\right)_{t=0}\ =\ 
\left(\frac{d\sigma^{pp}_{sd}} 
{dt}\right)_{t=0}^2\ .
\label{310}
 \eeq
 However, even within the Regge model this relation is broken due to
presence of Regge cuts. The usual accuracy of relations based on Regge
factorization is $10-20\%$. Besides, neither perturbative QCD calculations
\cite{bfkl}, nor phenomenological dipole cross sections fitted to DIS data
\cite{gbw} confirm Regge factorization. The very $Q^2$ dependence of the
Pomeron intercept observed at HERA is a direct evidence of lack of
factorization.

Unfortunately, the phenomenological dipole cross section $\sq$ which is
averaged over the size of the target proton, is not sufficient for calculating
double diffraction. Therefore, we will employ the approximate relation
(\ref{310}) in what follows, hoping that the corrections are not much larger
than in other known cases.

For the integrated double diffractive cross section relation (\ref{310})
can be rewritten as,
 \beq
\sigma^{pp}_{dd} =
\frac{(\sigma^{pp}_{sd})^2} 
{\sigma^{pp}_{el}}\ 
\frac{(B^{pp}_{sd})^2}{B^{pp}_{el}\,B^{pp}_{dd}}
\label{311}
 \eeq
 At the energy of HERA-B the elastic slope $B^{pp}_{el}=12.6\GeV^{-2}$;
$\sigma^{pp}_{el}=(\sigma^{pp}_{tot})^2/(16\pi B^{pp}_{el}) = 6.7\mb$;
$\sigma^{pp}_{sd}=3.5\mb$. The slope of the $\Pom\Pom\Reg$ part of single
diffraction relevant for Eq.~(\ref{311}), varies with Feynman $x_F$,
$B^{pp}_{sd} = B_0-2\alpha_\Pom^\prime\,\ln(1-x_F)$, where
$\alpha_\Pom^\prime=0.25\GeV^{-2}$, $B_0\approx 2\GeV^{-2}$ \cite{3R}. The
mass distribution of diffractive excitation of the valence quark skeleton of
the proton strongly peaks in the resonance region at $\overline{M}\approx
1.5\GeV$, which corresponds to $1-x_F = \overline{M}\,^2/s$. Then
$B^{pp}_{sd}=5.3\GeV^{-2}$. As for the double diffractive slope, it contains
only the Pomeron contribution, $B^{pp}_{dd}=
-2\alpha_\Pom^\prime\,\ln(1-x_F)=3.3\GeV^{-2}$. Eventually, we arrive at an
estimate for the double diffractive cross section at the energy of HERA-B,
$\sigma_{dd}^{pp} = 1.25\mb$. The same Eq.~(\ref{311}) leads to the double
diffraction cross sections $\sigma_{dd}^{pp} = 1.18\mb$ and $\sigma_{dd}^{pp}
= 0.47\mb$ at the energies of heavy ions at RHIC, $\sqrt{s}=0.2\TeV$, and at
LHC, $\sqrt{s}=5.5\TeV$, respectively.

\section{\boldmath$pA$ collisions:  Glauber model}\label{glauber}

The $pA$ elastic amplitude at impact parameter $b$ has the eikonal 
form \cite{glauber},
 \beq
 \Gamma^{pA}(\vec b;\{\vec s_j,z_j\}) =
1 - \prod_{k=1}^A\left[1-
 \Gamma^{pN}(\vec b-\vec s_k)\right]\ ,
 \label{10}
 \eeq
 where $\{\vec s_j,z_j\}$ denote the coordinates of an $i$-th target
nucleon; $i\Gamma^{pN}$ is the elastic scattering amplitude on a nucleon
normalized as,
 \beq
\st = 2\int d^2b\,\Re\Gamma^{pN}(b)\ .
\label{20}
 \eeq
 
 In the approximation of single particle nuclear density\footnote{We 
ignore the effect of motion of the center of gravity assuming the 
nucleus to be sufficiently heavy.},
 \beq
\rho_A(\vec b_1,z_1) = 
\int\prod_{i=2}^A d^3r_i\,
|\Psi_A(\{\vec r_j\})|^2\ ,
\label{40}
 \eeq
 the matrix element between the nuclear ground states reads,
 \beqn
&& \left\la0\Bigl|\Gamma^{pA}(\vec b;\{\vec s_j,z_j\})
\Bigr|0\right\ra 
\nonumber\\ &=&
1-\left[1-{1\over A}\int d^2s
\Gamma^{pN}(s)
\int\limits_{-\infty}^\infty dz
\rho_A(\vec b-\vec s,z)\right]^A.
\label{30}
 \eeqn
 Correspondingly, the total $pA$ cross section has the form,
 \beqn
&&\sta = 
2\Re\int d^2b\,\Biggl\{1 -
\Bigl[1 
\nonumber\\&-& \left.\left.
{1\over A}\int d^2s\,
\Gamma^{pN}(s)\,T_A(\vec b-\vec s)\right]^A\right\}
\approx
2\int d^2b\, 
\nonumber\\ &\times&
\left\{1-
\exp\left[-{1\over2}\,\st\,(1-i\alpha_{pp})\,
T^h_A(b)\right]\right\}\ ,
\label{50}
 \eeqn
 where $\alpha_{pp}$ is the ratio of the real to imaginary parts
of the forward $pp$ elastic amplitude;
 \beq
 T^h_A(b)= \frac{2}{\st}\int d^2s\, 
\Re\Gamma^{pN}(s)\,T_A(\vec b-\vec s)\ ;
\label{51} 
 \eeq
 and
 \beq
T_A(b) = \int_{-\infty}^\infty dz\,\rho_A(b,z)\ ,
\label{52}
 \eeq
 is the nuclear thickness function. 
 We use the Gaussian form of $\Gamma^{pN}(s)$,
 \beq
\Re \Gamma^{pN}(s) =
\frac{\st}{4\pi B_{el}^{pN}}\,
\exp\left(\frac{-s^2}{2B_{el}^{pN}}\right)\ .
\label{53}
 \eeq Note that the accuracy of the optical (exponential)  approximation
in (\ref{50}) is quite good, $\sim 10^{-3}$ for heavy nuclei, but for
numerical calculations we relay on the exact Glauber expressions
throughout the paper. In what follows we neglect the real part of the
elastic amplitude which gives quite a small correction, $\sim
\rho_{pp}^2/A^{2/3}$, and can be easily implemented.

We performed numerical calculations at $E_{lab}=920\GeV$, having in mind the
needs of the experiment HERA-B at DESY. At this energy we used
$\sigma^{pp}_{tot}=41.2\mb$, $B_{pp}=12.6\GeV^{-2}$ and
$\sigma^{pp}_{el}=(\sigma^{pp}_{tot})^2/16\pi B_{pp}=6.7\mb$. The Wood-Saxon
form was used for the nuclear density with parameters fixed by data on
electron-nucleus elastic scattering \cite{jager}, except carbon whose density
was taken in an oscillatory form \cite{sick},
 \beq
\rho_C(r)= \left(\frac{2a}{\pi}\right)^{{3\over2}}
\left(1+{4\over3}\,a\,r^2\right)\,e^{-a\,r^2}\ .
\label{56}
 \eeq
 The parameter $a$ was fitted to data for electron-carbon scattering,
$a=0.0143\GeV^2$, and we assumed identical distributions for protons and
neutrons. The results are depicted in Table~\ref{hera1}.

We also performed calculations at the energies of RHIC and LHC,
$\sqrt{s}=5.5\TeV$ using the input cross sections listed at the end of
Sect.~\ref{dd}. The results are presented in Tables~\ref{rhic} and \ref{lhc}
respectively.

\subsection{Elastic cross section}

The simplest process with a large rapidity gap (LRG) is elastic
scattering. It worth noting, however, that this channel is enhanced by 
absorptive corrections, while other LRG processes considered below are 
suppressed by these corrections. 

The elastic cross section according to (\ref{30}) reads,
 \beq
\sela=\int d^2b\, \left|1-
\exp\left[-{1\over2}\,
\st\,T^h_A(b)\right]\right|^2\ .
\label{60}
 \eeq
Numerical results are shown in Table~\ref{hera1}.

\subsection{Diffractive excitation of the beam and target}\label{pA-XA}

Diffractive excitation of the beam and/or target is another example of a
LRG process. Unfortunately, the Glauber model is a single channel
approximation and cannot treat properly diffractive excitation of the
beam. Nevertheless, the cross section of diffractive excitation of the
nucleus can be calculated, provided that the elastic and single
diffractive cross sections for $NN$ collisions are known.

If the nucleus is excited without new particle production, i.e. it breaks
up to nucleons and nuclear fragments, such a process, $pA\to pA^*$, is
called {\it quasielastic scattering}. Summing over final states of the
nucleus $|F\ra$ and applying the condition of completeness, one gets,
 \beqn
&&\sigma^{pA}_{qel}(pA\to pA^*) = \sum\limits_F
\int d^2b\ \left[\left\la0\left|
\Gamma^{pA}(b)\right|F\right\ra^\dagger
\right.
\nonumber\\ &\times& 
\left.
\left\la F\bigl|\Gamma^{pA}(b)\bigr|0\right\ra  -
 \left |\left\la 0\bigl|\Gamma^{pA}(b)
\bigr|0\right\ra\right|^2\right]\nonumber\\
&=& 
\int d^2b\left[\left\la0\left|
\bigl|\Gamma^{pA}(b)\bigr|^2
\right|0\right\ra
- \bigl|\left\la 0\left|\Gamma^{pA}(b)
\right|0\right\ra\Bigr|^2\right].
\label{70}
 \eeqn
 Here the cross section of elastic $pA\to pA$ scattering, Eq.~(\ref{60}), 
is subtracted.

The fist term in this expression contains the quadratic term $\int d^2s\, 
T_A(\vec b-\vec s)\left[\Gamma^{pN}(s)\right]^2 = T^h_A(b)\sel$.
Then the quasielastic cross section gets the form,
 \beqn
&& \sigma^{pA}_{qel}(pA\to pA^*) 
\nonumber \\ &=& 
\int d^2b\,\left\{
\exp\left[-\sinhad\,T^h_A(b)\right]-
\exp\left[-\st\,T^h_A(b)\right]\right\}
\nonumber \\ &\approx& 
\seln \int d^2b\,T_A(b)\,
\exp\left[-\sinhad\,T^h_A(b)\right]\ .
\label{90}
 \eeqn

Another possibility to excite the nucleus is to excite a bound nucleon,
$pA\to pY$. To specify the terminology, following \cite{carvalho} we call
this channel target single diffraction (tsd). Since the nucleus breaks up
anyway, and the debris of the bound nucleon stay in the target
fragmentation region, they cannot fill the rapidity gap, therefore their
final state interaction do not affect the LRG cross section. Then it must
have the same form as the quasielastic cross section Eq.~(\ref{90}),
except the normalization factor,
 \beq
\sigma^{pA}_{tsd}(pA\to pY)= 
\frac{\sigma^{pp}_{sd}}{\sigma^{pp}_{el}}\,
\sigma^{pA}_{qel}(pA\to pA^*)\ .
\label{100}
 \eeq

We fix the single diffractive $pp$ cross section at
$\sigma^{pp}_{sd}=3.5\mb$ for $E_{lab}=920\GeV$ in accordance with
\cite{dino}. For the energies of RHIC and LHC we extrapolate the Tevatron
data \cite{dino} assuming no energy independence, $\sigma^{pp}_{sd}=4\mb$.

The numerical results for the target dissociation cross sections are 
shown in Tables~\ref{rhic} and \ref{lhc}.

\section{Color transparency}\label{ct}

The light-cone dipole representation was proposed in \cite{zkl} as an
effective tool for calculation of hadronic cross sections and nuclear
shadowing, relying on the observations that color dipoles are the
eigenstates of hadronic interactions at high energies, and the eigenstate 
method \cite{kl1} is a powerful tool for summing up all Gribov inelastic 
corrections.

The key ingredient of this approach, the cross section of the
dipole-nucleon interaction, $\sq(r_T)$, is an universal and flavor
independent function which depends only on transverse separation $r_T$ and
energy. Applications of the dipole formalism to nuclei are especially
simple, if the energy is sufficiently high to freeze the fluctuations of
the dipole size during its propagation through the nucleus.
Otherwise one should rely on the light-cone Green function technique
\cite{kz91,krt1,krt2}, which takes care of these fluctuations.

Due to color screening colorless point-like dipoles cannot interact with an
external color field. Since the underlying theory is non-abelian, the
interaction cross section for such dipoles vanishes at $r_T\to 0$ as
$\sq(r_T)\propto r_T^2$ \cite{zkl}, the phenomenon called color
transparency\footnote{Actually, the cross section behaves as $r_T^2\ln(r_T^2)$
\cite{zkl}, but with a good accuracy one can fix the logarithm at an effective
separation typical for the process under consideration.}. At high energies
nuclei are transparent for small-size fluctuations of the incoming hadron,
therefore the exponential attenuation suggested by the eikonal Glauber formula
cannot be correct and should underestimate the nuclear transparency, i.e.
overestimate the total hadron-nucleus cross section.

Thus the Glauber approach is subject to modifications called Gribov's
corrections.  Originally those corrections were proposed in hadronic
representation where they look like intermediate diffractive excitations
\cite{gribov}. The lowest order correction is expressed via the single
diffraction cross section $\sigma_{sd}^{hN}(hN\to XN)$ measured
experimentally \cite{gribov,kk},
 \beqn
\Delta\sigma^{hA}_{tot} &=&
- 4\pi\int d^2b
\exp\left[-{1\over2}\,\sigma^{hN}_{tot}\,
T_A(b)\right]
\nonumber\\ &\times&
\int\limits_{M_{min}^2} dM^2\,
\left.\frac{d\sigma_{sd}^{hN}}
{dM^2\,dp_T^2}\right|_{p_T=0}
\int\limits_{-\infty}^{\infty}dz_1\,
\rho_A(b,z_1)
\nonumber\\ &\times&
\int\limits_{z_1}^{\infty}dz_2\,
\rho_A(b,z_1)\,e^{iq_L(z_2-z_1)}\ ,
\label{110}
 \eeqn
 where 
 \beq
q_L=\frac{M^2-m_h^2}{2E_h}\ .
\label{120}
\eeq
 Therefore, one might think that this is a model-independent calculation. 
However, the cross section of interaction of the intermediate inelastic 
state $X$ is unknown and is assumed to be equal to $\sigma^{hN}_{tot}$.

Formally, all the observables calculated either in color-dipole, or
hadronic representations must be identical. Nevertheless, although the
correction Eq.~(\ref{110}) is negative, i.e. it makes the nuclear matter
more transparent, it cannot reproduce the effect of color transparency
\cite{kn} which results from compensation of many diagonal (positive) and
off-diagonal (negative) amplitudes. Indeed, for heavy nuclei this
correction vanishes exponentially with nuclear thickness $T_A(b)$.

The dipole representation allows to sum up the inelastic corrections to all
orders \cite{zkl}. For a dipole of a fixed size $r_T$ the eikonal form is
exact, since the dipole is the eigenstate of interaction. Therefore nuclear
transparency, which is the no-interaction probability of propagation of a
$\bar qq$ dipole with fixed separation $\vec r_T$ through nuclear medium,
reads,
 \beq
Tr(r_T)=e^{-\sq(r_T)\,T_A}\ .
\label{128}
 \eeq
 In averaging over dipole
sizes important contribution comes only from small $\sq(r_T)\lsim 1/T_A$.
Therefore, for a sufficiently long path in the nuclear medium only very 
small values of $r_T$ contribute, and any model for 
the dipole cross section must have the same behavior $\sq(r_T)\propto 
r_T^2$. Then \cite{zkl},
 \beq
Tr=\left\la e^{-\sq(r_T)\,T_A}\right\ra 
\propto \frac{1}{T_A}\ .
\label{130}
 \eeq
 This result should be compared with the exponential attenuation in the
Glauber model, $Tr = \exp(-\st\,T_A)$. Such a difference cannot result from
the lowest order inelastic correction Eq.~(\ref{110})  which has the same
exponential dependence on $T_A$. This is a full color transparency effect
which must include all the higher order inelastic corrections. It is
characterized by the nuclear saturation scale $Q_s^2\propto T_A$ and can be
treated perturbatively for sufficiently heavy nuclei or at very high energies.
In reality, such large nuclear thicknesses are beyond the reach of existing
nuclei. For this reason, one should not rely on the limiting behavior
Eq.~(\ref{130}), but employ realistic models for the dipole cross section. In
what follows we demonstrate how important is this fact.

\section{\boldmath$pA$ collisions:  the total, elastic and single
diffractive cross sections}\label{coherent}

\subsection{Excitation of the valence quark skeleton of the 
proton}\label{3q}

Once models for the proton wave function and the dipole cross section
are chosen one is in a position to perform calculations for Gribov's
corrections to different nuclear reactions, in particular LRG processes.
The total and elastic cross sections read,
 \beqn
&& \sigma_{tot}^{pA} =
2\int d^2b\,\left[1-
\left\la e^{-{1\over2}\,
\sigma_{3q}(r_i)\,\T}\right\ra\right] 
\nonumber \\ &\equiv&
2\int d^2b
\left[1-
\int \prod\limits_{i=1}^3 d^2r_i\,
\left|\Psi_N(r_j)\right|^2\,
\right.\nonumber \\ &\times&
e^{-{1\over2}\,
\sigma_{3q}(\vec r_k)\,T_A(b)}\Biggr]\ ;
\label{312}
 \eeqn
 \beq
\sigma_{el}^{pA} = \int d^2b\,\left[1-
\left\la e^{-{1\over2}\,
\sigma_{3q}(r_i)\,\T}\right\ra\right]^2
\label{314}
 \eeq

Single diffractive excitation of the projectile proton, $pA\to XA$ cannot
be treated properly within the single-channel Glauber model approximation
which assumes that the projectile hadron is an eigenstate of the
interaction. To generalize the model one should introduce off-diagonal
diffractive amplitudes, but then one faces the same problem as in the case
of higher order Gribov's corrections: lack of experimental information for
those amplitudes. And the remedy is the same, to switch to the dipole
representation.

The cross section of single diffraction on a nucleus related to excitation of
the valence quark skeleton, $[\sigma_{sd}]_{\Pom\Pom\Reg}$, is given by
Eq.~(\ref{165}), but with the replacement, $\sq\Rightarrow \sqa = 2\int
d^2b\,\left\{1- \exp[-{1\over2}\sq T_A(b)\right\}$.

 \beqn
&& \int dM_X^2\,\left[
\frac{d\sigma_{sd}(pA\to XA)}{dM_X^2} 
\right]_{\Pom\Pom\Reg} 
\nonumber\\ &=&
\int d^2b\,\left[
\left\la e^{-\sigma_{3q}(r_i)\,\T}\right\ra
\right.\nonumber\\ &-&\left. 
\left\la e^{-{1\over2}\,
\sigma_{3q}(r_i)\,\T}\right\ra^2\right]
\label{318}
 \eeqn
 Both terms in this expression are vanishingly small for heavy nuclei,
except at the very periphery. Therefore, the cross section of single
diffraction is expected to rise as $A^{1/3}$, but the coefficient should
be sensitive to the inelastic shadowing corrections.

Although comparison with single diffraction performed in previous sections
for four different variants of the dipole model clearly demonstrated that
only the last two, which employ the saturated dipole cross section, may
be realistic, we will try all four cases for nuclear reactions to get an
idea of the theoretical uncertainties.

The inelastic corrections to $\sta$ have been well detected experimentally
\cite{murthy,gsponer} and found to be rather small, about $10\%$. However,
one should not think that inelastic shadowing is always a weak effect. In
fact, for heavy nuclei it affects dramatically the exponential term in
(\ref{50}), but the term itself is very small compared to one, since the
unitarity limit is almost saturated. However, LRG channels (except elastic
scattering), e.g. the single diffraction Eq.~(\ref{318}), may be very
sensitive to these corrections, since their cross section is proportional
to nuclear transparency.

It was demonstrated in Sect.~\ref{ct} that for hadrons heavy nuclei are much 
more
transparent than the Glauber model predicts, due to color
transparency and the presence of small-size dipoles in the hadronic wave
function. As a result, the exponential attenuation switches to an inverse
linear dependence on $T_A$, Eq.~(\ref{130}). This is, however, an
asymptotic behavior valid in the limit of $\T\to\infty$, since for real
nuclei the result depends on the actual modeling of $\Psi_N(\vec r_i)$ and
$\sq(r_T)$.

\noi{\bf Model I}\\
 Here we employ the quark-diquark model Eq.~(\ref{210}) for the proton
wave function and the dipole cross section $\sq(r_T)\propto r_T^2$. Then
the averaged exponential terms in (\ref{312})-(\ref{318}) read,
 \beq
\left\la\exp\left[-{1\over2}\,
\sq(r_T)\,\T\right]\right\ra_{I} 
=\frac{1}{1+{1\over2}\,\stn\,\T}
\label{330}
 \eeq

The results of computing the total and elastic $pA$ cross sections are
compared with the Glauber model in Table~\ref{hera1}. The effect of inelastic
shadowing is rather large, in fact the total cross section is reduced by about
$20\%$. Although no data are available at this energy, extrapolation from
lower energies \cite{murthy} is hardly compatible with such a correction.
Apparently this is closely related to the found overestimation of diffraction
by this model of the dipole cross section.

\noi{\bf Model II}\\
Since the probability to be found in a point like 
configuration is less for three- than two-body system,
one should expect more opaque nuclei in this variant.
Indeed,
 \beqn
&& \left\la\exp\left[-{1\over2}\,
\sigma_{3q}(r_i)\T\right]\right\ra_{II}
\nonumber\\ &=&
\int \prod\limits_i^3 d^2r_i\, 
\left|\Psi_N(\vec r_1,\vec r_2,\vec
r_3)\right|^2\, 
e^{-{1\over2}\,
\sigma_{3q}(\vec r_1,\vec r_2,\vec
r_3)\, \T} \nonumber\\ &=& 
\frac{1}{\left[1+{1\over4}\,
\stn\,\T\right]^2}\ .
 \label{340}
 \eeqn
 In this case the nuclear transparency is quadratic, rather than linear
function of the inverse nuclear thickness \cite{mine}.

The results of the calculation, depicted in Table~\ref{hera1}, show that the
inelastic correction is about $10\%$ of the total cross section for heavy
nuclei. This looks much better than for the previous model, and is compatible
with what may be expected as an extrapolation of data at lower energies.  
However, it is too early to jump to conclusions: the triple-Pomeron part of
shadowing has not been included yet.

 Naturally, the elastic cross section follows the total cross section and is
reduced by the inelastic corrections as well. As we mentioned, it may be at
variance with other LRG channels which one may expect to be enhanced, if
nuclei are more transparent. However, further calculations show that the
situation is more complicated.

\noi{\bf Model III}\\
The steep rise with $r_T$ of the dipole cross section, $\sq(r_T)\propto
r_T^2$, used above is justified only for small, but not for large $r_T$.
This is why it overestimates diffraction.  More reliable calculations can
be done using a more realistic phenomenological cross section which levels 
off
at large $r_T$, as described in Sect.~\ref{xsect}.

We can perform analytic calculations with the saturated cross section
Eq.~(\ref{180}) expanding the Glauber exponentials. Then the mean value of the
exponential terms in Eqs.~(\ref{312})-(\ref{318}) for the total cross section
gets the form,
 \beqn
&& \left\la\exp\left[-{1\over2}\,
\sq(r_T)\,\T\right]\right\ra_{III} 
\nonumber\\ &=&
e^{-\frac{1}{2}\,
\s0\,T^h_A(b)}\,
\sum\limits_{n=0}^\infty\,
\frac{[\s0\,T^h_A(b)]^n}
{2^n\,(1+n\,\delta)\,n!}\ ,
\label{370}
 \eeqn
 where $\sigma_0(s)$ and $\delta$ are defined in (\ref{255}).

As could be expected, the numerical results shown in Table~\ref{hera1} 
demonstrate a
weaker effect of inelastic shadowing compared to the previous variants.

\noi{\bf Model IV}\\
This case involves the symmetric 3q-wave function of the proton and the 
saturated cross section Eq.~(\ref{180}). Expanding again the Glauber 
exponential and performing part of the integrations analytically, we arrive 
at the following result,
 \beqn
&& \left\la\exp\left[-{1\over2}\,
\sigma_{3q}(r_i)\T\right]\right\ra_{IV} 
= {3\over4}\,e^{-{3\over4}\,
\s0\,T^h_A(b)}
\nonumber\\ &\times&
\int\limits_0^\infty d\xi
\left[\sum\limits_{n=0}^\infty\,
e^{-\frac{\xi}{4(1+n\gamma)}}\ 
\frac{[\s0\,T^h_A(b)]^n}
{4^{n}\,(1+n\,\gamma)\,n!}\,
\right]^3\ ,
\label{380}
 \eeqn
 where $\gamma$ is defined in (\ref{300}).

As far as the Glauber exponentials are averaged over the eigenstates of
interaction, for four different models, Eqs.~(\ref{330})-(\ref{380}), we are
in a position to calculate the total, elastic and inelastic $pA$ cross
sections given by Eqs.~(\ref{312}), (\ref{314}) and (\ref{318})  
respectively.

The numerical results for four nuclei and energy $\sqrt{s}=41.6\GeV$ of
the experiment HERA-B are displayed in Table~\ref{hera1} in parentheses
for four different combinations of the models for the valence quark wave
function of the proton and the dipole cross section. 
 \begin{center}
 \begin{table}[thb]
\Caption{
 \label{hera1} Cross sections of coherent reactions calculated within the
Glauber model and also corrected for inelastic shadowing which either
excludes (in parentheses), or involves the corrections for gluon
shadowing.} 
\begin{tabular}{|c|c|c|c|c|c|c|}
 \hline
 \vphantom{\bigg\vert}
   Obs. &Nucl.
  & Glauber
  & Model                   
  & Model                   
  & Model                   
  & Model   \\
[-0.4cm]
&&model& I & II & III & IV \\
\hline &&&&&&\\[-6mm]
&W & 3073  & (2462) & (2727) 
& (2908) & (3000) \\
 [-0.2cm]
&&& 2382 & 2641 & 2831 &2918 \\
$\sigma^{pA}_{tot}
$&Ti & 1159 & (938)  & (1028) 
& (1103) & (1130) \\
 [-0.2cm]
$[\mb]$
&&& 915 & 1003 & 1077 & 1102 \\
&Al& 726  & (594) & (647) 
& (692) & (707) \\
[-0.2cm]
&&& 582 & 632 & 677 & 690\\
&C & 380 & (349) & (372) 
& (363) & (369) \\
[-0.2cm]
&&& 344 & 366 & 357 & 361\\
\hline &&&&&&\\[-6mm]
&W& 1196  & (748) & (931) 
& (1061) & (1137) \\
[-0.2cm]
&&& 695 & 867 & 1001 & 1071 \\
$\sigma^{pA}_{el}
$ & Ti & 352  & (217) & (268) 
& (313) & (331) \\
[-0.2cm]
$[\mb]$
&&& 205 & 252 & 296 & 312 \\
&Al& 196  & (122) & (149) 
& (174) & (183) \\
[-0.2cm]
&&& 116 & 141 & 165 & 173 \\
&C& 79.1  & ( 51.6) & (61.5) 
& (70.5) & (73.3) \\
[-0.2cm]
&&& 49.6 & 58.9 & 67.4 & 70.0\\
\hline &&&&&&\\[-6mm]
&W & - & (153) & (82.3) 
& (57.2) & (19.2) \\
[-0.2cm]
&&& 156 & 86.5 & 59.0 & 20.6 \\
$\sigma^{pA}_{sd}$
&Ti& - & (66.3) & (39.4) 
& (23.5) & (9.4) \\
[-0.2cm]
$_{(\Pom\Pom\Reg)}$
&&& 65.7 & 39.5 & 23.1 & 9.4\\
$[\mb]$
&Al& -  & (42.4) & (25.7) 
& (14.4) & (6.0) \\
[-0.2cm]
&&& 41.8 & 25.5 & 14.0 & 6.0\\
&C & - & (22.8) & (14.0) 
& (6.9) & (3.1) \\
[-0.2cm]
&&& 22.4 & 13.8 & 6.7 & 3.0\\
\hline &&&&&&\\[-6mm]
&W & 1.2 & (4.1) & (2.8) 
& (2.1) & (1.6) \\
[-0.2cm]
&&& 3.7 & 2.7 & 2.0 & 1.6 \\
$\sigma^{pA}_{sd}$
&Ti& 0.6 & (1.0) & (0.8) 
& (0.7) & (0.6) \\
[-0.2cm]
$_{(3\Pom})$
&&& 0.9 & 0.8 & 0.6 & 0.6\\
$[\mb]$
&Al& 0.3 & (0.5) & (0.4) 
& (0.4) & (0.4) \\
[-0.2cm]
&&& 0.5 & 0.4 & 0.4 & 0.4\\
&C & 0.1 & (0.2) & (0.2) 
& (0.2) & (0.2) \\
[-0.2cm]
&&& 0.2 & 0.2 & 0.2 & 0.2\\
\hline   
\end{tabular}
 \end{table}
\end{center}
 Our predictions for $p-Au$ collisions at RHIC and for $p-Pb$ collisions
at LHC are depicted in Tables~\ref{rhic} and \ref{lhc} respectively.

 As one could have expected, the nonrealistic models I and II, which grossly
overestimate single diffraction, also overestimate the inelastic corrections.
On the other hand, the more realistic models III and IV lead to quite moderate
deviation from the Glauber model expectations. However, this is not the end of
the story, since the inelastic shadowing related gluonic excitations is still
to be included.

Note that the result for single diffraction in Tables~\ref{hera1}, \ref{rhic},
\ref{lhc} demonstrate much higher sensitivity to inelastic corrections than
for elastic scattering. This is easy to understand, in the black disc limit
the elastic cross section reaches its maximum, $\pi R_A^2$, and cannot be
varied by any corrections. At the same time, diffraction vanishes, therefore
is extremely sensitive to the transparency of the nucleus, it is maximal for
model I, but minimal for IV.

\subsection{Diffractive excitation of the gluonic degrees of freedom}
\label{sdg}

The higher Fock components of the proton wave function contain more partons,
gluons, sea quarks. Apparently, the more degrees of freedom, the larger are
the Gribov corrections. In particular, diffractive gluon radiation,
corresponding to the triple-Pomeron term, makes nuclear matter more
transparent, i.e. leads to a further reduction of the total $pA$ cross
section. These corrections will be taken into account in Sect.~\ref{gluons}.

Diffractive gluon radiation also contributes to the single diffractive
process $pA\to XA$ and creates the triple-Pomeron tail $1/M^2$ in the
effective mass distribution. Correspondingly, the single-diffraction cross
section Eq.~(\ref{318}) must be corrected for this excitation channel. The
cross section of coherent gluon radiation on a nucleus is given by a
straightforward generalization of Eq.~(\ref{303}),
 \beqn   
&&\left[\frac{d\sigma(pA\to XA)}{dx_F}\right]_{3\Pom}
\nonumber\\ &=&
\frac{3}{4\pi(1-x_F)}
\int d^2b\,\left\la e^{-{1\over2}\,
\sigma_{3q}(r_i,s)\,\T}\right\ra^2
\nonumber\\ &\times&
\int d^2r_T\,\biggl|\Psi_{qG}(\vec r_T)\biggr|^2
\left[1-e^{-{1\over2}\,
\widetilde\sigma(r_T,s)T_A(b)}\right]^2,
\label{390}
 \eeqn
 with the same notations. Here the first factor implies the absorptive
corrections, analogous to those in (\ref{306}). They are, however, much
stronger than in $pp$ collisions and almost terminate diffraction on heavy
nuclei, except at the very periphery.  This factor is averaged over the proton
size weighted with its wave function squared, as in (\ref{166}). Therefore it
depends on the models for the proton wave function considered above. Since we
employ the phenomenological cross section fitted to data, the possibility of
gluon radiation during propagation of the proton though the nucleus is
included.

 The integral over $r_T$ in (\ref{390}) takes into account the diffractive
channels containing gluons coherently radiated by a valence quark propagating
through a nucleus of thickness $T_A(b)$.  We make use of the smallness of the
mean quark-gluon separation and add up incoherently the gluons radiated by
different projectile valence quarks. Besides, the two models for the dipole
cross section under discussion are almost identical at such small separations,
$r_0\sim 0.3\fm$.
 \begin{widetext}
\begin{center}
\begin{table}[h]
\Caption{
 \label{rhic}
 Predictions for RHIC for different LRG channels. All the cross sections
are for proton-gold collisions and are in $\mb$.  The cross sections shown
in parentheses are corrected for inelastic shadowing related only to
valence quark fluctuations, while the bottom numbers are also corrected
for gluon shadowing. The cross sections which turned out to be smaller
than the numerical accuracy of the calculations are put equal to zero.}
 \begin{tabular}{|c|c|c|c|c|c|c|c|c|}
 \hline
 \vphantom{\bigg\vert}
   Model & 
$\sigma^{pPb}_{tot}$
  & $\sigma^{pPb}_{el}$
  & $\left[\sigma^{pPb}_{sd}\right]_{\Pom\Pom\Reg}$
  & $\left[\sigma^{pPb}_{sd}\right]_{3\Pom}$
  & $\sigma^{pPb}_{qel}$
  & $\sigma^{pPb}_{qsd}$
  & $\sigma^{pPb}_{tsd}$
  &$\sigma^{pPb}_{dd}$
   \\
\hline &&&&&&&&\\[-6mm]
Glauber& 3616.8 & 1446.8 & - & 5.1 & 98.6 & - & 42.3 & - \\
\hline &&&&&&&&\\[-6mm]
III& (3524.4)  & (1367.6) & (31.8) & (7.3)  
& (95.8) & (3.1) & (41.2)& (3.1)\\
[-0.2cm]
& 3457.5& 1313.8& 33.3 & 7.6 & 96.2 & 3.1 & 41.4 & 3.1\\
\hline &&&&&&&&\\[-6mm]
IV & (3582.0) & (1417.7) & (8.1) & (5.8) 
& (98.4) & (0.6)& (42.3)&(0.6)\\
[-0.2cm]
& 3514.1 & 1362.3 & 8.9 & 6.3 & 98.9 & 0.6 & 42.53 &0.6\\
\hline   
\end{tabular}
 \end{table}

 \begin{table}[h]
 \Caption{
 \label{lhc}
 The same as in Table~\ref{rhic}, but for proton-lead collisions at 
LHC.}
\begin{tabular}{|c|c|c|c|c|c|c|c|c|}
 \hline
 \vphantom{\bigg\vert}
   Model & 
$\sigma^{pPb}_{tot}$
  & $\sigma^{pPb}_{el}$
  & $\left[\sigma^{pPb}_{sd}\right]_{\Pom\Pom\Reg}$
  & $\left[\sigma^{pPb}_{sd}\right]_{3\Pom}$
  & $\sigma^{pPb}_{qel}$
  & $\sigma^{pPb}_{qsd}$
  & $\sigma^{pPb}_{tsd}$
  &$\sigma^{pPb}_{dd}$
   \\
\hline &&&&&&&&\\[-6mm]
Glauber& 4241.5 & 1794.9 & - & 28.8 & 141.43 & - & 22.9 & - \\
\hline &&&&&&&&\\[-6mm]
III& (4222.9)  & (1778.7) & (5.5) & 31.8  
& (142.1) & (0.0)& (23.0)& (0.0)\\
[-0.2cm]
& 4194.2& 1755.6 & 5.8 & 33.4 & 141.8 & 0.0 & 23.0 & 0.0\\
\hline &&&&&&&&\\[-6mm]
IV & (4235.2) & (1789.9) & (0.8) & 29.5 
& (142.8) & (0.0)& (23.1)&(0.0)\\
[-0.2cm]
&4207.1 & 1767.3 & 0.9 & 31.2 & 142.5 & 0.0 & 23.1 &0.0\\
\hline   
\end{tabular}
 \end{table}
\end{center}
 \end{widetext}
Therefore for the sake of simplicity we use the simple quadratic form,
$\sigma_{\bar qq}\propto r_T^2$. The accuracy of this approximation is
greatly enhanced by the cut off imposed on large separations by the first
exponential factor in Eq.~(\ref{390}), which gets the form,
 \beqn   
&&\left[\frac{d\sigma(pA\to XA)}{dx_F}\right]_{3\Pom}
= \frac{\alpha_s}{\pi^2(1-x_F)}
\int d^2b
\nonumber\\ &\times&
\left\la e^{-{1\over2}\,
\sigma_{3q}(r_i)\,\T}\right\ra^2\,
\ln\left[1+\frac{\epsilon_A^2(b,s)}
{1+2\epsilon_A(b,s)}\right],
\label{395}
 \eeqn
 where
 \beq
\epsilon_A(b,s)=\frac{9\,r_0^2}
{16\,R_0^2(s)}\,K(s)\,\sigma_0(s)\,T_A(b)\ .
\label{397}
 \eeq

The absorptive corrections, given in (\ref{395}) by the exponential
factor, are factorized from the coherent diffractive gluon radiation,
given by the logarithmic factor. For heavy nuclei these absorptive
corrections are much stronger than for $pp$ collisions (see
Sect.~\ref{unitarity}). They practically eliminate diffraction on nuclear 
targets except the very periphery. Therefore, the cross section 
Eq.~(\ref{397}) varies as $A^{1/3}$.

Since the main contribution comes from peripheral collisions where
the projectile proton finds very few nucleons, the absorptive 
corrections for $NN$ scattering considered earlier in Sect.~\ref{unitarity} 
may be important and have to be included into (\ref{397}). This is done in 
the same way as in Sect.~\ref{unitarity}, namely using the relation 
Eq.~(\ref{306}) which results in the suppression factor $K(s)$ in 
Eq.~(\ref{397}),
 \beq
K(s)=1-\frac{1}{4\pi}\,
\frac{\sigma^{pp}_{tot}(s)}
{B^{pp}_{el}(s)+B^{pp}_{3\Pom}}\ .
\label{399}
 \eeq
 Note that although this factor decreases with energy, it always remains
positive. Indeed, $K>0$ even in the Froissart limit, where \cite{dklt}
$\sigma_{tot}=4\pi\alpha_{\Pom}^{\prime}\,\epsilon\ln^2(s/M_0^2)$ and
$B_{el}=\alpha_{\Pom}^{\prime}\,\epsilon\ln^2(s/M_0^2)$, where $\epsilon
= \alpha_\Pom(0)-1\approx 0.08$. The triple Pomeron slope
$B^{pp}_{3\Pom}$ depends on $x_F$ according to (\ref{304}). We evaluate
it at the mean value of $\overline{x}_F$. The magnitude of
$B^{pp}_{3\Pom}(\overline{x}_F)$ varies very slowly with energy, as a
double logarithm. It rises only by $10\%$ between the energies of HERA-B
and LHC.

Only the absorptive correction factor in (\ref{397}) depends on the model
for the proton wave function and on the shape of the dipole cross section.  
It can be evaluated either within the Glauber approximation, or including
the inelastic shadowing corrections calculated in accordance with either
of the four models considered above. We fix $\alpha_s=0.4$ as was explained
in Sect.~\ref{unitarity} and integrate over $x_F$ from 0.9 up to
$1-M_0^2/s$. The results are depicted in Table~\ref{hera1} for the energy
of HERA-B, and in Tables~\ref{rhic} and \ref{lhc} for the energies of RHIC
and LHC respectively.

The coherent triple-Pomeron diffraction on nuclei at the energy of HERA-B
turns out to be amazingly small. This can be understood as follows. At this
energy the value of $\epsilon(s)$ in (\ref{305}) is sufficiently small to
expand the logarithm up to the first order of $\epsilon^2$. The same is true
for Eq.~(\ref{397}) provided that the absorptive corrections eliminate the
contribution of large $T_A(b)$. Then, comparing the cross section
Eq.~(\ref{305}) integrated over $p_T$, and the one presented in
Eq.~(\ref{397}), we see that they are related via the replacement,
 \beq
\frac{1}{B^{pp}_{3\Pom}} \Rightarrow
A\la T_A\ra\ ,
\label{400}
 \eeq
 where the mean nuclear thickness for this process is given by
 \beq   
\la T_A\ra = \frac{1}{A}
\int d^2b\,\left\la e^{-{1\over2}\,
\sigma_{3q}(r_i)\,\T}\right\ra^2\,
T_A^2(b)\ .
\label{410}
 \eeq
 This can be estimated as,
 \beq
\la T_A\ra\ \approx 
\frac{2\pi w R_A}{A(\sigma^{pp}_{tot})^2}\ ,
\label{420}
 \eeq
 where $R_A\approx 1.12\fm\times A^{1/3}$ and $w\sim 0.5\fm$ are the nuclear
radius and edge thickness for the Woods-Saxon nuclear density \cite{jager}.
Thus, the replacement Eq.~(\ref{400})  leads to a reduction of the cross
section by about a factor of five for heavy nuclei, $A\sim 200$, and much more
for light nuclei. This explains the smallness of the coherent diffractive
gluon radiation by extremely strong absorptive corrections. In the black disk
limit with vanishing edge thickness ($w\to 0$) no diffractive gluon radiation
is possible.

\section{Quasielastic scattering/excitation of the 
projectile, \boldmath$pA\to p(X)A^*$}\label{incoherent}

The simplest channel of nuclear excitation is a quasielastic breakup of
the nucleus, caused by elastic scattering, $pN\to pN$, of the projectile
proton on one of the bound nucleons. In this case the nucleus breaks up
into fragments without particle production. It is difficult to control
this condition experimentally, but is easy to calculate it.

One should modify Eq.~(\ref{70}), first sum the nuclear excitations $A^*$
and integrate over impact parameter $\vec s$, then average over the quark
positions $\vec r_i$ and $\vec r^\prime_j$ in each of the two amplitudes,
 \beqn
&& \sigma^{pA}_{qel}(pA\to pA^*)
\nonumber\\&=&
\int d^2b
\left\{\sum\limits_F \left\la 
0\left|\Gamma^{pA}(b,\vec r_i)
\right|F\right\ra^\dagger_{s,r_i}
\left\la F\left|\Gamma^{pA}(b,\vec 
r_i)\right|0\right\ra_{s,r_j^\prime}
\right.\nonumber\\ &-& \left.
\left\la 0\left|
\Gamma^{pA}(b,\vec r_i)
\right|0\right\ra_{s,r_i}^\dagger
\left\la 0\left|
\Gamma^{pA}(b,\vec r_j^\prime)
\right|0\right\ra_{s,r_j^\prime}\right\}
\nonumber\\&=&
\int d^2b\left\{\left\la0\left|
\left[\Gamma^{pA}(b,\vec r_i)\right]^\dagger
\Gamma^{pA}(b,\vec r_j^\prime)
\right|0\right\ra_{s,r_i,r^\prime_j}
\right. \nonumber\\&-& \left.
\left\la 0\left|
\Gamma^{pA}(b,\vec r_i)
\right|0\right\ra_{s,r_i}^2\right\}\ .
\label{440}
 \eeqn
 Here $\vec s$ is the impact parameter of the projectile proton relative to
the bound nucleons, and $\vec r_i$ are the relative transverse positions of
the valence quarks in the projectile proton. After integration over impact
parameter $\vec s$ [see in (\ref{70})] we get,
 \beqn
\sigma^{pA}_{qel} &=&
\int d^2b \left\la\left\la\exp\left[
- {1\over2}\sigma(\vec r_i)\,\T 
- {1\over2}\sigma(\vec r_j^\prime)\,\T\right]
\right.\right.
\nonumber\\ &\times& 
\left.\left.\left\{
\exp\left[\frac{\sigma(\vec r_i)\,\sigma(\vec r_j^\prime)}
{16\pi B_{el}}\,\T\right]\ -1\,\right\}
\right\ra_{r_i}
\right\ra_{r_j^\prime}
\nonumber\\ &=&
\int d^2b\,
\sum\limits_{k=1}\,\frac{1}{k!}\,
\left[\frac{\T}{4\pi B_{el}}\right]^k
\nonumber\\ &\times& 
\left\{\frac{\partial^k}{\partial(T_A)^k}\,
\left\la 
\exp\left[- {1\over2}\,\sigma(\vec r_i)\,\T\right]
\right\ra_{r_i}\right\}^2\ .
\label{450}
 \eeqn
 This series quickly converges due to smallness of the elastic cross 
section. Even the fist term provides a reasonable accuracy. We control 
the accuracy to be within $1\%$.

 If we sum over all excitations of the proton containing no radiated
gluons and apply the condition of completeness, a delta-function
$\delta(\vec r_i - \vec r_i^\prime)$ emerges, leading to the following
expression for the sum of quasielastic and quasi-single-diffractive
channels,
 \beqn
&& \sigma^{pA}_{qel}(pA\to pA^*) +
\sigma^{pA}_{qsd}(pA\to XA^*) 
\nonumber\\ &=& 
\int d^2b\, \Bigl\la\exp\left[
- \sigma(\vec r_i)\T\right]
 \nonumber\\ &\times& \left. 
\left\{\exp\left[\frac{\sigma^2(\vec r_i)}
{16\pi B_{el}}\T\right] -1\right\}
\right\ra_{r_i}
\nonumber\\ &=&
\int d^2b\,
\sum\limits_{k=1}\,\frac{1}{k!}\,
\left[\frac{\T}{16\pi B_{el}}\right]^k\ 
\nonumber\\ &\times& 
\frac{\partial^{2k}}{\partial(T_A)^{2k}}\,
\left\la 
\exp\left[-\sigma(\vec r_i)\,\T\right]
\right\ra_{r_i}.
\label{460}
 \eeqn
 
With these expressions, Eqs.~(\ref{450})-(\ref{460}), we can calculate
the quasielastic and quasidiffractive cross sections employing different 
models for the dipole cross section and the proton wave function.

\noi{\bf Model I}\\
In this simplest case of the mesonic type wave function of the proton and
$\sigma(\vec r_i)\equiv\sq(r_T)\propto r_T^2$, the quasielastic and 
quasi-diffractive cross sections
read respectively,
 \beq
\sigma^{pA}_{qel} =
\sum\limits_{k=1}\,k!
\int d^2b\,
\frac{\left[\sel\,\T\right]^k}
{\left[1+{1\over2}\,\st\,\T\right]^{2k+2}}.
\label{470}
 \eeq

 \beq
\sigma^{pA}_{qel} +
\sigma^{pA}_{qsd} =
\sum\limits_{k=1}\,
\frac{(2k)!}{k!}\int d^2b\,
\frac{\left[\sel\,\T\right]^k}
{\left[1+\st\,\T\right]^{2k+1}}.
\label{480}
 \eeq

\noi{\bf Model II}\\
In the case of a symmetric valence quark wave function and dipole cross 
section $\sigma(\vec r_i)\equiv\sigma_{3q}(\vec r_i)$ the
quasielastic cross section takes the form,
 \beq
\sigma^{pA}_{qel} =
\sum\limits_{k=1}\frac{(k+1)(k+1)!}{2^{2k}}
\int d^2b
\frac{\left[\sel\,\T\right]^k}
{\left[1+{1\over4}\st\T\right]^{2k+4}}.
\label{490}
 \eeq

 This result may look surprising. Indeed, the quasielastic cross section
Eq.~(\ref{90}) is proportional to nuclear transparency, i.e. the survival
probability of a proton propagating through the nucleus. That is given by
the mean value Eq.~(\ref{340}) squared, i.e. the fourth power of $T_A$ in
the denominator. That would mean more nuclear transparency and larger
quasielastic cross section compared to the Glauber model. Eq.~(\ref{490}),
however, has the leading term which behaves as the inverse sixth power of
$T_A$. Although at large $T_A$ the exponential Glauber transparency is
less than any power of $T_A$, it turns out that for real nuclei inelastic
shadowing makes nuclei less transparent, at variance with the simplified
expectation. The reason why the nucleus is less transparent than is
suggested by Eq.~(\ref{340}) is an extra weight factor $\sq(r)$ in the
quasielastic amplitude. This factor suppresses small-size projectile
components for which the nucleus is transparent.

Using completeness we can also calculate the sum of the cross sections of
quasielastic and quasidiffractive scattering, the result reads,
 \beq
\sigma^{pA}_{qel} +
\sigma^{pA}_{qsd} =
\sum\limits_{k=1}
\frac{(2k+1)!}{2^{2k}\,k!}\int d^2b
\frac{\left[\sel\,\T\right]^k}
{\left[1+{1\over2}\st\T\right]^{2k+2}}.
\label{500}
 \eeq
 In this case the leading term behaves like $T_A^{-3}$ at large $T_A$, 
i.e. very heavy nuclei are much more transparent for quasidiffractive, 
than quasielastic processes.

\noi{\bf Models III and IV}\\
 We skip the cumbersome expressions for the quasielastic and
quasi-diffractive cross sections in this case. We use instead equations
(\ref{450}) and (\ref{460}) respectively and perform calculations
numerically. For the averaged Glauber exponential $\left\la \exp\left[-
{1\over2}\sigma(\vec r_i)\,\T\right] \right\ra_{r_i}$ we employ
Eqs.~(\ref{370}) and (\ref{380}) for models III and IV respectively.

So far we calculated the quasi-diffractive cross section related to proton
excitations without gluon radiation. Now we should include diffractive
gluon bremsstrahlung, as it is done for coherent diffraction in
Sect.~\ref{sdg}.  This can be done via a simple replacement in the above
equations (\ref{450})-(\ref{500}),
 \beq
\sigma^{pA}_{qsd}\,
\Rightarrow\,
\frac{\sigma^{pp}_{sd}}
{\left[\sigma^{pp}_{sd}\right]_{\Pom\Pom\Reg}}\,
\sigma^{pA}_{qsd}\ .
\label{505}
 \eeq

 The results for $\sigma_{qel}^{pA}$ and $\sigma_{qsd}^{pA}$ are depicted
in parentheses in Table~\ref{hera2} for different nuclei at the energy
of HERA-B, and in Tables~\ref{rhic} and \ref{lhc} for gold at RHIC and lead
at LHC respectively. In both models III and IV the quasi-diffractive and
double-diffractive cross sections are very small due to very low nuclear
transparency close to the black disc limit, so small at the energy of LHC
that in model IV we could not reach a sufficient numerical accuracy.
Therefore these cross sections are set to zero in Table~\ref{lhc}.
 \begin{table}[thb]
\Caption{
 \label{hera2} Same as in table~\ref{hera1}, but involving
diffractive break up of the nucleus.}
 \begin{center}
\begin{tabular}{|c|c|c|c|c|c|c|}
 \hline
 \vphantom{\bigg\vert}
   Obs. &Nucl.
  & Glauber
  & Model
  & Model
  & Model
  & Model
\\[-0.4cm]
&&model& I&II&III&IV \\
\hline &&&&&&\\[-6mm]
&W & 88.2 & (59.1) & (45.9) & (73.0) & (77.3) \\
[-0.2cm]
&&&         59.0 & 46.0 & 73.5 & 78.0\\
$\sigma^{pA}_{qel}[\mb]$ 
&Ti & 63.9 & (42.5) & (35.3) & (53.2) & (56.0) \\ 
[-0.2cm]
&&&        42.0 & 35.2 & 52.8 & 55.5\\
&Al& 48.8 & (32.8) & (30.0) & (40.7) & (42.7) \\
[-0.2cm]
&&&         32.4 & 27.8 & 40.3 & 42.2\\
&C & 34.5 & (23.2) & (20.5) & (28.6) & (29.7) \\
[-0.2cm]
&&&         22.8 & 20.3 & 28.1 & 29.2\\
\hline &&&&&&\\[-6mm]
& W & - & (41.9) & (22.6)  & (6.2) & (2.5) \\
[-0.2cm]
&&&        40.7 & 22.3 & 6.1 & 2.5\\ 
$\sigma^{pA}_{qsd}[\mb]$ 
&Ti& - & (31.4) & (18.7) & (4.4) & (1.8) \\
[-0.2cm]
&&&       30.6 & 18.6 & 4.5 & 1.9\\ 
&Al& -  & (24.8) & (15.5) & (3.5) & (1.5) \\
[-0.2cm]
&&&        24.2 & 15.4 & 3.6 & 1.5 \\
&C& - & (18.2) & (11.6) & (2.7) & (1.2) \\
[-0.2cm]
&&&      17.7 & 11.6 & 2.7 & 1.2 \\ 
\hline &&&&&&\\[-6mm]
& W & 42.1 & (28.2) & (21.9) & (34.8) & (36.9) \\ 
[-0.2cm]
&&&           28.1 & 22.1 & 35.1 & 37.2 \\ 
$\sigma^{pA}_{tsd}[\mb]$ 
&Ti& 30.5 & (20.3) & (16.9) & (25.4) & (26.7) \\
[-0.2cm]
&&&          20.1 & 16.8 & 25.2 & 26.5 \\ 
&Al& 23.3  & (15.7) & (13.3) & (19.4) & (20.4) \\
[-0.2cm]
&&&           15.5 & 13.3 & 19.2 & 20.1 \\
&C& 16.5 & (11.0) & (9.8) & (13.6) & (14.2) \\
[-0.2cm]
&&&          10.9 & 9.7 & 13.4 & 14.0 \\ 
\hline &&&&&&\\[-6mm]
&W & - & (15.0) & (8.1) & (2.2) & (0.9) \\
[-0.2cm]
&&&       14.5 & 8.0 & 2.2 & 0.9 \\
$\sigma^{pA}_{dd}[\mb]$&Ti
& -     & (11.2) & (6.7) & (1.6) & (0.7) \\
[-0.2cm]
&&&       10.8 & 6.7 & 1.6 & 0.7\\
&Al& -  & (8.9) & (5.5) & (1.3) & (0.5) \\
[-0.2cm]
&&&        8.6 & 5.5 & 1.3 & 0.5 \\
&C & - & (6.5) & (4.3) & (1.0) & (0.4) \\
[-0.2cm]
&&&       6.3 & 4.1 & 1.0 & 0.4 \\
\hline   
\end{tabular}
\end{center}
 \end{table}

 One can notice a much higher sensitivity to the models for
quasidiffractive scattering compared to quasielastic. In the former case
this simply reflects the tremendous model dependence of single diffraction,
pointed out in Sect.~\ref{diff}. In the latter case there is a partial
compensation between the steep rise of the dipole cross section with the
dipole size ($\propto r_T^4$ at small $r_T$) and the exponential fall of
nuclear transparency.

\section{Diffractive excitation of bound nucleons}\label{tdd}

Another LRG process, not explored so far, is diffractive excitations of bound
nucleons in the target. It may happen along with or without excitation of the
beam. We call these cases target single diffraction, or double diffraction 
respectively.

Since the phenomenological dipole cross is averaged over the target proton
wave function, it is insufficient for these reactions which need knowledge
of a double-dipole cross section. If, however, one assumes that in $pp$
collisions the dependence on the beam dipole size is the same for
reactions with or without target excitation, then one can relate the cross
sections in question to known quasielastic and quasidiffractive ones,
 \beqn
\sigma^{pA}_{tsd}(pA\to pY) &=& 
\frac{\sigma^{pp}_{sd}}{\sigma^{pp}_{el}}\,
\sigma^{pA}_{qel}(pA\to pA^*)\ ;
\label{510}\\
\sigma^{pA}_{dd}(pA\to XY) &=& 
\frac{\sigma^{pp}_{dd}}{\sigma^{pp}_{sd}}\,
\sigma^{pA}_{qsd}(pA\to XA^*)\ ;
\label{520}
 \eeqn
 Numerical results are depicted in Table~\ref{hera2} for HERA-B and in
Tables~\ref{rhic} and \ref{lhc} for RHIC and LHC respectively.

\section{Gluon shadowing and the triple-Pomeron diffraction}\label{gluons}

Eikonalization of the lowest Fock state $|3q\ra$ of the proton done in
(\ref{312}) corresponds to the Bethe-Heitler regime of gluon radiation.
Indeed, gluon bremsstrahlung is responsible for the rising energy dependence
of the phenomenological cross section (\ref{170}), and in the eikonal form
(\ref{10}) one assumes that the whole spectrum of gluons is multiply radiated.
However, the Landau-Pomeranchuk-Migdal (LPM) effect \cite{lp,m} is known to
suppress radiation in multiple interactions. Since the main part of the
inelastic cross section at high energies is related to gluon radiation, the
LPM effect becomes a suppression of the cross section. This is a
quantum-mechanical interference phenomenon and it is a part of the suppression
called Gribov's inelastic shadowing. The way how it is taken into account in
the QCD dipole picture is by the inclusion of higher Fock states, $|3qG\ra$,
etc. Each of these states represents a colorless dipole and its elastic
amplitude on a nucleon is subject to eikonalization.

As we already mentioned, the eikonalization procedure requires the
fluctuation lifetime to be much longer than the nuclear size. Otherwise,
one has to take into account the "breathing" of the fluctuation during
propagation through a nucleus, which can be done by applying the
light-cone Green function technique \cite{kz91,kst1,kst2}. In hadronic
representation this is equivalent to saying that all the longitudinal
momenta transfers must be much smaller than the mean free path of the
hadron in the nucleus. Otherwise, the phase shifts and interferences are
to be included \cite{kl2}.

The c.m. energies of HERA-B, RHIC and LHC are sufficiently high to treat
the lowest Fock state containing only valence quarks as "frozen" by the
Lorentz time dilation during propagation through the nucleus.  Indeed, for
the excitations with the typical nucleon resonance masses, the coherence
length is sufficiently long compared to the nuclear size. This is why we
applied eikonalization without hesitation so far. Such an approximation,
however, never works for the higher Fock states containing gluons. Indeed,
since the gluon is a vector, the integration over effective mass of the
fluctuation is divergent, $dM^2/M^2$, which is the standard triple-Pomeron
behavior. Therefore, the colliding energy never can be sufficiently high
to neglect the large-mass tail. For this reason the inelastic shadowing
corrections related to excitation gluonic degrees of freedom and
calculated in the tree-approximation never saturate, and keep rising
logarithmically with energy.

There are, however non-linear effects which are expected to stop the rise
of inelastic corrections at high energies. This is related to the
phenomenon of gluon saturation \cite{glr,al} or color glass condensate
\cite{mv}. The strength of these nonlinear effect is model dependent. It
is a rather mild effect within the model of small gluonic spots explained
in Sect.~\ref{3-pom}. The reason is simple, in spite of a sufficient
longitudinal overlap of gluon clouds originated from different nucleons,
there is insufficient overlap in the transverse plane. This fact leads to
a delay of the onset of saturation up to very high energies, since the
transverse radius squared of the gluonic clouds rise with energy very
slowly, logarithmically, with a small coefficient of the order of
$0.1\GeV^{-2}$

The details of the calculation of inelastic corrections related to excitation
of gluonic degrees of freedom can be found in \cite{kst2,mine}. The results
for nuclear cross sections corrected for gluon shadowing are shown in
Tables~\ref{hera1}, \ref{rhic}, \ref{lhc}, and \ref{hera2}.

\section{Conclusions}\label{conclusions}

In this paper we study the dynamics of LRG processes in $pp$ and $pA$
collisions basing on the light-cone dipole approach which allows to
develop phenomenology not only for hard, but also for soft reactions.

One of our objectives is an improvement of accuracy of calculations for
nuclear effects in LRG processes. First of all, we tested some popular models
for the dipole-proton cross section and the proton wave function on their
ability to reproduce the cross section of diffractive excitation of the
valence quark skeleton of the proton. Data show that the forward diffractive
cross section is amazingly small, only about $10\%$ of the elastic one. We
conclude that the models based on the dipole cross section which quadratically
depends on quark separation fail badly to explain even the order of magnitude
of the single diffractive $pp$ cross section. This model, however, is quite
popular in the literature devoted to nuclear effects, in particular color
transparency, since it helps to simplify the calculations. Apparently, those
results cannot be realistic. On the other hand, we found the saturated shape
of the dipole cross section which levels off at large separations to be quite
successful in explaining the data on $pp$ diffraction.

As for nuclear effects, the popular Glauber model cannot treat properly most
of the off-diagonal diffractive processes, since this is a single-channel
approximation.  Based on the color-dipole representation we develop techniques
for calculating the cross sections of LRG processes, both diagonal and
off-diagonal. This method allows to sum the Gribov inelastic shadowing
corrections to all orders.  These corrections make nuclear matter more
transparent and reduce the total hadron-nucleus cross section. At the same
time, their influence on other diffractive reactions depends on a complicated
color transparency interplay making nuclei more transparent for small size
hadronic fluctuations, but simultaneously suppressing the strength of the
interaction with bound nucleons, due to the same effect. This is confirmed by
our numerical results for the cross sections of a variety of channels
presented in Tables~\ref{hera1}-\ref{hera2}. We found that models I and II
based on the quadratic dependence of the cross section on the dipole size
grossly overestimate the Gribov corrections compared to more realistic
variants III and IV, based on the saturated form of the dipole cross section.

Available data for the Gribov corrections \cite{murthy, gsponer} show that
they rise with energy, what results from the falling energy dependence of the
longitudinal momentum transfer. At sufficiently high energies the
$\Pom\Pom\Reg$ part of the single diffraction saturates, since the
longitudinal momentum transfer vanishes, while the triple-Pomeron part keeps
rising logarithmically.  Comparing, however, the results depicted in
Tables~\ref{hera1} and \ref{lhc} we see that the inelastic corrections, i.e.
the deviation from the Glauber model, vary from $5-7\%$ at the energy of
HERA-B down to about $1-2\%$ at LHC. Such a dramatic reduction signals the
closeness of the unitarity limit. Indeed, in this regime, also called
black-disk limit, different Fock components of the proton interact with the
same cross section. Therefore the incoming and outgoing waves consist of the
same superposition of Fock states, then only elastic scattering is possible.  
Similar suppression of diffraction is expected for $pp$ scattering when it
reaches the Froissart regime, $\sigma^{pp}_{sd}/\sigma^{pp}_{tot}\propto
1/\ln(s)$. However, the onset of this effect on nuclear targets can be
observed at much lower energies.

 \begin{acknowledgments}

We are thankful to Bernhard Schwingenheuer who has inspired us for this
study, Joao Carvalho, Werner Hoffman, Michael Schmelling, and Joachim
Spengler, for constant interest and critical discussions. Special thanks go
to Hans-J\"urgen Pirner for reading the manuscript and making valuable
improving suggestions.  Work was supported in part by Fondecyt (Chile)
grants 1030355, 1050519 and 1050589, and by DFG (Germany) grant PI182/3-1.
Part of this work was completed while B.K. was employed by the
Max-Planck-Institut f\"ur Kernphysik, Heidelberg.

\end{acknowledgments}


\begin{thebibliography}{99}

\bibitem{kst2} B.Z.~Kopeliovich, A.~Sch\"afer and A.V.~Tarasov, Phys.
Rev. {\bf D62} (2000) 054022.

\bibitem{k3p} B.Z.~Kopeliovich, I.K.~Potashnikova, B.~Povh, E.~Predazzi,
Phys. Rev. Lett. {\bf 85}, 507 (2000); Phys. Rev. {\bf D63}, 0540012001
(2002).

\bibitem{qm04} B.Z.~Kopeliovich and B.~Povh, J. Phys. G{\bf 30}, S999 
(2004).

\bibitem{shuryak} E.~Shuryak and I.~Zahed Phys. Rev. D {\bf 69}  
014011 (2004).
 
\bibitem{florian} D.~de~Florian and R.~Sassot, Phys. Rev. D {\bf 69},
074028 (2004).

\bibitem{dino} K.Goulianos, J.~Montanha, Phys. Rev. {\bf D59}, 114017
(1999)

\bibitem{peter} S.~Erhan and P.E.~Schlein, Phys. Lett. {\bf B427} (1998) 
389.

\bibitem{zkl} A.B.~Zamolodchikov, B.Z.~Kopeliovich and L.I.~Lapidus,
Sov. Phys. JETP Lett. {\bf 33} (1981) 612

\bibitem{ihkt} J.~H\"ufner, Yu.P.~Ivanov, B.Z.~Kopeliovich and
A.V.~Tarasov, Phys. Rev. {\bf D62} (2000) 094022; Phys. Rev.{\bf C66}
(2002) 024903.

\bibitem{low} F.E.~Low, Phys. Rev. D {\bf 12}, 163 (1975). 

\bibitem{nussinov} S.~Nussinov, Phys. Rev. D {\bf 14}, 246 (1976).

\bibitem{kp} B.Z.~Kopeliovich and B.~Povh, Z. Phys. {\bf A356} (1997) 467.

\bibitem{pion} S.~Amendolia et al., Nucl. Phys. {\bf B277} (1986) 186.

\bibitem{gbw} K.~Golec-Biernat and M.~W\"usthoff, Phys. Rev. D{\bf 59}
(1999) 014017.

\bibitem{sgbk} A.M.~Stasto, K.~Golec-Biernat, J.~Kwiecinski, Phys. Rev.
Lett. {\bf 86}, 596 (2001).

\bibitem{mine} B.Z.~Kopeliovich, Phys. Rev. {\bf C68} (2003) 044906.

\bibitem{sb} I.A.~Schmidt and R.~Blankenbecler, Phys. Rev. D {\bf 15}, 
3321 (1977).

\bibitem{diquark} M.~Anselmino, E.~ Predazzi, S.~Ekelin, S.~Fredriksson,
and D.B. Lichtenberg, Rev. Mod. Phys. {\bf 65} (1993) 1199.

\bibitem{kz} B.Z.~Kopeliovich and B.G.~Zakharov, Phys. Lett.  {\bf 211B}
(1988) 221; Sov.  J.  Nucl. Phys. {\bf 48} (1988) 136; {\it ibid} {\bf 49}
(1989) 674; Z. Phys. {\bf C43} (1989) 241; Sov. Phys. Particles and
Nuclei, {\bf 22} (1991) 140

\bibitem{3R} Yu.M.~Kazarinov, B.Z.~Kopeliovich, L.I.~Lapidus and 
I.K.~Potashnikova, JETP {\bf 70} (1976) 1152.

\bibitem{km} Yu.V.~Kovchegov and A.H.~Mueller, Nucl.  Phys. B {\bf 529},
451 (1998).

\bibitem{kkt} D.~Kharzeev, Y.V.~Kovchegov and K.~Tuchin, Phys.  Lett. B
{\bf 599}, 23 (2004).

\bibitem{r-ch} R. Rosenfelder, Phys. Lett. B {\bf 479}, 381 (2000). 

\bibitem{amaldi} U.~Amaldi and K.R.~Schubert, Nucl. Phys. B{\bf 166}, 301 
(1980).

\bibitem{kp1} B.Z.~Kopeliovich, B.~Povh and E.~Predazzi, 
Phys. Lett. B{\bf 405}, 361 (1997).

\bibitem{glm} E. Gotsman, E. Levin, U. Maor, Phys. Lett. B 
{\bf 438}, 229 (1998). 

\bibitem{pdg} Review of Particle Physics,
D.E. Groom et al, Eur. Phys. J. C{\bf 15}, 1 (2000).

\bibitem{bfkl} E.A.~Kuraev, L.N.~Lipatov and V.S.~Fadin, Sov.  Phys.  
JETP {\bf 44} (1976) 443 ; {\bf 45} (1977) 199; Ya.Ya.~Balitskii and
L.I.~Lipatov, Sov.  J.  Nucl. Phys. {\bf 28} (1978) 822; L.N.~Lipatov,
Sov. Phys. JETP {\bf 63} (1986) 904.

\bibitem{glauber} R.J.~Glauber, Phys. Rev. {\bf 100} (1955) 42.

\bibitem{jager} C.W.~De~Jager, H.~De~Vries, C.~De~Vries,
Atom. Data Nucl. Data Tables {\bf 36} (1987) 495.

\bibitem{sick} I.~Sick, J.S.~Mccarthy, Nucl. Phys. {\bf A150} 
(1970) 631.

\bibitem{carvalho} J. Carvalho, Nucl. Phys. {\bf A725} (2003)  269.

\bibitem{kl1} B.Z.~Kopeliovich and L.I.~Lapidus, Sov. Phys.
JETP Lett. {\bf 28} (1978) 664

\bibitem{kz91} B.Z.~Kopeliovich and B.G.~Zakharov, Phys. Rev. {\bf D44}
(1991) 3466.

\bibitem{krt1} B.Z.~Kopeliovich, J.~Raufeisen and A.V.~Tarasov,
Phys. Lett. {\bf B440} (1998) 151.
 
\bibitem{krt2}  B.Z.~Kopeliovich, J.~Raufeisen and A.V.~Tarasov,
Phys. Rev. {\bf C62} (2000) 035204.

\bibitem{gribov} V.N.~Gribov, Sov. Phys. JETP {\bf 56} (1968) 892.

\bibitem{kk} V.~Karmanov and L.A.~Kondratyuk, Sov. Phys. JETP
Lett. {\bf 18} (1973) 266.

\bibitem{kn} B.Z.~Kopeliovich and J.~Nemchik, Phys. Lett. {\bf B368} (1996)
187

\bibitem{murthy} P.V.R.~Murthy et al., Nucl. Phys. {\bf B92}, 269 (1975).

\bibitem{gsponer} A. Gsponer et al., Phys. Rev. Lett. {\bf 42}, 9 (1979). 

\bibitem{dklt} M.I.~Dubovikov, B.Z.~Kopeliovich, L.I.~Lapidus
K.A.~Ter-Martirosyan, Nucl. Phys. {\bf B123} 147, (1977).

\bibitem{lp} L.D.~Landau and I.Ya.~Pomeranchuk, {\it ZhETF} {\bf 24}, 505
(1953); \\ L.D.~Landau, I.Ya.~Pomeranchuk, {\it Doklady AN SSSR} {\bf
92}, 735 (1953); \\ E.L.~Feinberg and I.Ya.~Pomeranchuk, {\it Doklady AN
SSSR} {\bf 93}, 439 (1953); \\ I.Ya.~Pomeranchuk, {\it Doklady AN SSSR}
{\bf 96}, 265 (1954); \\ I.Ya.~Pomeranchuk, {\it Doklady AN SSSR} {\bf
96}, 481 (1954): \\ E.L.~Feinberg, I.Ya.~Pomeranchuk, {\it Nuovo Cim.
Suppl.} {\bf 4}, 652 (1956).

\bibitem{m} A.B.~Migdal, Phys. Rev. {\bf 103}, 1811 (1956).

\bibitem{kst1} B.Z.~Kopeliovich, A.~Sch\"afer and A.V.~Tarasov,
Phys. Rev. {\bf C59},1609 (1999).

\bibitem{kl2} B.Z.~Kopeliovich and L.I.~Lapidus, Sov. Phys. JETP Lett. 
{\bf 32}, 612 (1980).

\bibitem{glr} L.V.~Gribov, E.M.~Levin and M.G.~Ryskin, Nucl. Phys. {\bf
B188} (1981) 555; Phys. Rep. {\bf 100} (1983) 1.

\bibitem{al} A.H.~Mueller, Eur. Phys. J. A {\bf 1}, 19 (1998).

\bibitem{mv} L.~McLerran and R.~Venugopalan, Phys. Rev. D {\bf 49}, 2233
(1994); {\bf 49}, 3352 (1994); {\bf 49}, 2225 (1994).

\end{thebibliography}
\end{document}